\documentclass[aps,prb,twocolumn,superscriptaddress,10pt]{revtex4-2} 
\pdfoutput=1 
\usepackage[utf8]{inputenc}
\usepackage[dvips]{graphicx}
\usepackage{bm,mathtools,bbold,physics}
\usepackage{braket}
\usepackage{amsmath,amsfonts,amsthm,amssymb}
\usepackage[bookmarksnumbered,pdfpagelabels=true,plainpages=false,colorlinks=true,linkcolor=blue,citecolor=red,urlcolor=blue]{hyperref}
\usepackage[capitalise]{cleveref}
\usepackage{tabularx}

\def\ri{\mathrm i}
\def\re{\mathrm e}

\def\rd{\mathrm d}

\def\B{\mathcal B}

\def\NB{\mathcal N_{\mathcal B}}

\def\PB{\hat P_{\mathcal B}}
\def\QB{\hat Q_{\mathcal B}}

\def\H{\hat H}
\def\HB{\hat H_{\mathcal B}}
\def\Hbloch{\hat H_0}
\def\Hpert{U(\CanM,\R)}

\def\Hpertpara{U_{\rm para}(\CanM,\R)}
\def\Hpertdia{U_{\rm dia}(\R)}
\def\Hpertphi{U_{\phi}(\R)}

\def\E{\mathcal E}

\def\OB{\hat{O}_{\mathcal B}}

\newcommand{\proj}[1]{\left( #1 \right)_\mathcal{B}}
\newcommand{\projs}[1]{( #1 )_\mathcal{B}}

\def\R{\hat{\bm r}}
\def\RB{\hat{\bm r}_{\mathcal B}}
\def\dtRB{\dot{\hat{\bm r}}_{\mathcal B}}
\newcommand{\RBc}[1]{{\hat r_{#1,\mathcal B}}}

\def\dtRinterB{\dot{\hat{\bm r}}_{\perp}}

\def\CrM{\hat{\bm k}}

\def\dtCrMB{\dot{\hat{\bm k}}_{\mathcal B}}

\def\CanM{\hat{\bm p}}
\newcommand{\CanMc}[1]{{\hat p_{#1}}}

\def\BCebf{{\bm{\mathrm A}}}
\newcommand{\BC}[1]{{\hat{\mathrm A}_{#1}}}
\newcommand{\BCe}[1]{{\mathrm A_{#1}}}
\newcommand{\BCB}[1]{{\hat{\mathrm A}_{#1,\mathcal B}}}
\begin{document}

\title{Quantum geometry in the dynamics of band-projected operators}

\author{Chen Xu}
% \email{chen.xu@uni.lu}
 \affiliation{Department of Physics and Materials Science, University of Luxembourg, 1511 Luxembourg, Luxembourg}
 \affiliation{Institute of Theoretical Physics and W\"urzburg-Dresden Cluster of Excellence ct.qmat,\\ Technische Universit\"at Dresden, 01069 Dresden, Germany}

\author{Andreas Haller}
% \email{andreas.haller@uni.lu}
\affiliation{Department of Physics and Materials Science, University of Luxembourg, 1511 Luxembourg, Luxembourg}

\author{Suraj Hegde}
 % \email{suraj.hegde@tu-dresden.de}
 \affiliation{Institute of Theoretical Physics and W\"urzburg-Dresden Cluster of Excellence ct.qmat,\\ Technische Universit\"at Dresden, 01069 Dresden, Germany}
\affiliation{Indian Institute of Science Education and Research Thiruvananthapuram, Vithura, 695551, India}

 \author{Tobias Meng}
 % \email{tobias.meng@tu-dresden.de}
\affiliation{Institute of Theoretical Physics and W\"urzburg-Dresden Cluster of Excellence ct.qmat,\\ Technische Universit\"at Dresden, 01069 Dresden, Germany}

 \author{Thomas L. Schmidt}
\email{thomas.schmidt@uni.lu}
 \affiliation{Department of Physics and Materials Science, University of Luxembourg, 1511 Luxembourg, Luxembourg}

\date{\today}

\begin{abstract}
We study the dynamics of electrons in crystalline solids in the presence of inhomogeneous external electric and magnetic fields. We present a manifestly gauge-invariant operator-based approach without relying on a semiclassical wavepacket construction, and derive the field-induced corrections to the equations of motion at the operator level. This includes the Berry curvature induced anomalous velocity and contributions arising from the quantum geometry of the Bloch bands. We show explicitly how these multi-band effects are manifested in an effective single band approximation. We present a formalism that allows for a systematic expansion to an arbitrary order in the inhomogeneity of the applied fields, as well as a way to compute the matrix elements in Bloch basis.
\end{abstract}

\maketitle

{\it Introduction.} One of the central results of solid-state theory is Bloch's theorem~\cite{Bloch1929berDQ}, which states that the energy spectrum of a crystalline system is organized into bands and that the states within each band can be labeled by a quantum number associated with the lattice translation symmetry: the crystal momentum $\bm{k}$. It was realized already in the early days that the geometry of the Bloch band structure gives rise to corrections to the group velocity $\partial \mathcal{E}/\partial\bm{k}$, and is responsible for interesting transport phenomena such as the intrinsic anomalous Hall effect in ferromagnets \cite{karplus,kohn,luttinger,Adams1959, Blount1962FormalismsOB, refId0, ahe,Haldane2004}. In fact, the authors of Refs.~\cite{karplus,Adams1959} already derived what is known today as the Berry connection~\cite{Berry:1984jv} and, within a semiclassical wavepacket approach, the anomalous velocity related to the Berry curvature \cite{changniu1995,sundarmniu,changniu}. With the advent of topological materials, such semiclassical dynamics has received renewed interest, and for example, is applied various novel transport properties of topological semimetals \cite{sonspivak,Hosur2013,mooreorenstein,orensteinmoore,Burkov2015,Li2016,Sekine2021,mirage}. 

The geometry underlying the projective Hilbert space of quantum systems, broadly termed quantum geometry, has recently come to the limelight in various contexts. The Berry curvature may be viewed as the imaginary component of a quantity called the quantum geometric tensor, and its real part, known as the quantum metric or Fubini-Study metric \cite{Provost1980,Bengtsson2006,Avdoshkin2023}, has been found to bring about a wide range of phenomena, for instance in fractional topological phases \cite{parameswaran,Parameswaran2013,roy,claassen,mera,meraozawa}, superfluidity and superconductivity in flat bands \cite{julkutorma,torma,Torma:2021pac,huhtinentorma}, and nonlinear optical responses \cite{ahnguonaogaosa,ahn,Ahn2020TheoryOO,hsuahn}. In particular, previous works have shown that under the influence of an inhomogeneous electric field, the quantum metric tensor as well as the associated Christoffel symbols induce further contributions to the electron dynamics in metallic systems \cite{Lapa2019,gaoxiao,Kozii2021}. 

The canonical way of obtaining the semiclassical equations of motion, which take into account the Berry curvature, is based on constructing a wavepacket using a basis that self-consistently depends on the instantaneous position of the center of mass position \cite{sundarmniu,changniu,SHINDOU2005399}. While straightforward, this approach obscures the physical origin of such corrections to the dynamics. It is also not manifest how the Ehrenfest theorem~\cite{Sakurai2020} is bypassed, which among other things predicts no correction to the position dynamics when the external potential field is only position-dependent. Furthermore, the use of instantaneous eigenstates seems to necessitate a solution the perturbed Hamiltonian, but in actual applications of the semiclassical equations of motion, the Berry curvature used is that of the free lattice Hamiltonian. In this Letter, we present an operator-based approach without reference to wavepackets, and demonstrate that such geometric contributions are fundamentally multi-band effects, which emerge when using an effective low-energy description of the electron dynamics within a restricted set of bands. 

{\it Model.} Throughout this work, we use natural units $\hbar=c=1$. We consider a fermion with charge $q$ and mass $m$ moving in a lattice potential under the influence of an applied static electromagnetic field. It is described by the Hamiltonian  
\begin{align}
    \H = \frac{\hat{\bm \pi}^2}{2m} + q\phi(\hat{\bm r}) + V(\hat{\bm r})
    =
    \Hbloch + \Hpert
    .
    \label{eq:free_hamiltonian}
\end{align}
Here, $\hat{\bm{\pi}}=\hat{\bm{p}}-q\bm{A}(\hat{\bm r})$ is the kinematic momentum, and $\Hbloch=\CanM^2/(2m) + V(\R) $ is the lattice Hamiltonian which by virtue of Bloch's theorem~\cite{Bloch1929berDQ, Ashcroft76} can be decomposed as $\Hbloch = \sum_{n\bm k}\E_{n\bm k}\ket{\psi_{n\bm k}}\bra{\psi_{n\bm k}}$, using the Bloch eigenstates and eigenenergies, $\Hbloch\ket{\psi_{n\bm k}}=\E_{n\bm k}\ket{\psi_{n\bm k}}$. The external perturbations are given by
\begin{align}
    \Hpert
    =
    \Hpertphi+\Hpertdia + \Hpertpara 
    .
\end{align}
The terms correspond, respectively, to the electrostatic potential $\Hpertphi = q\phi(\R)$, the diamagnetic terms $\Hpertdia = q^2 \bm A(\R)^2/(2m$), and the paramagnetic contribution $\Hpertpara = -q (\bm A(\R) \cdot \CanM+\CanM\cdot\bm A(\R))/(2m)$. 
Since we are interested in time-independent electromagnetic fields, we choose a static gauge for simplicity. The inhomogeneous external fields can then be written in operator form as
\begin{align}
    \bm B(\R)
    =
    \partial_{\R}\times \bm A(\R)
    ,\quad
    \bm E(\R)
    =
    -\partial_{\R} \phi(\R)
    .
\end{align}
%===========================================================================
%===========================================================================
{\it Projected dynamics.} We assume that the applied electromagnetic field has a long wavelength compared to the lattice spacing, such that Bloch's theorem remains applicable, and a weak field strength such that $\Hpert$ is small compared to any band gaps. In such a case, the low-energy dynamics may be described by restricting to a subset of bands near the Fermi energy. For example, under weak electric fields, only the bands near the Fermi level of a metal need to be considered for studying transport properties. However, such a band restriction has nontrivial effects. Roughly speaking, band-projected operators, which represent the effective observables in a restricted Hilbert space $\mathcal B\subset\mathcal H$, will have different commutation relations compared to their unprojected counterparts. For example, the projected position operators $\hat{r}_{j\mathcal{B}}$ ($j=x,y,z$) satisfy $[\hat{r}_{j\mathcal{B}},\hat{r}_{k\mathcal{B}}]=\ri\hat{\Omega}_{jk}$, where $\hat{\Omega}_{jk}$ is the operator form of Berry curvature (see Supplemental Material). As we shall demonstrate below, this is indeed the origin of quantum geometric contributions such as the anomalous velocity, which is related to the Berry curvature.
%
%
%
% 
% 

%===========================================================================
%===========================================================================
{\it Dynamics of band-projected operators.} We perform the projection to low-energy bands via the projector associated with a subset $\NB$ of bands,
\begin{align}
    \PB = \sum_{n\in\NB} \sum_{\bm k}\ket{\psi_{n\bm k}}\bra{\psi_{n\bm k}}
    ,\quad
    \QB = \mathbb1 - \PB
    .
\end{align}
Note that the projection operator $\PB$ and its complement $\QB$ are invariant under a band-dependent gauge transformation of the Bloch states $\ket{\psi_{n\bm{k}}}\rightarrow e^{\ri\phi_n(\bm{k})}\ket{\psi_{n\bm{k}}}$. We denote a band-projected operator as $\OB = \PB\hat O\PB$, and we consider the effective dynamics within the projected subspace as given by the band-projected temporal derivative \cite{refId0} 
\begin{align}
 \dot{\hat O}_\mathcal{B} \equiv \PB (\rd\OB/\rd t ) \PB,
\end{align}
where the total time derivative acts only on the operator $\hat O_\B$. Assuming no explicit time-dependence in $\hat{O}$, in the Heisenberg picture this is equivalent to, 
\begin{align}
    \dot{\hat{O}}_\mathcal{B}=-\ri[\hat{O}_{\mathcal{B}},
    \hat{H}_{\mathcal{B}}].\label{eq:projectedheisenberg}
\end{align}
Physically, this means that the projected dynamics of an operator in the projected subspace is governed by an effective Hamiltonian $\hat{H}_{\mathcal{B}}$. This may also be seen as a first-order effect in perturbation theory~\cite{nolting2018theoretical,refId0}. Note that both the observable and the Hamiltonian are projected, which is different from projecting only the commutator. To motivate this definition, consider the expectation value of the observable $\hat{O}$ evaluated in a state within the projected Hilbert space, $O_\B(t) =\bra{\psi_\B}\hat{O}_\B(t)\ket{\psi_\B}$. The dynamics of the observable is then given by $\ri\dot{O}_B(t)=\bra{\psi_\B}[\hat {O}_\B(t),\hat{H}]\ket{\psi_\B}=\bra{\psi_\B}[\hat {O}_\B(t),\hat{H}_\B]\ket{\psi_\B}$, in agreement with Eq.~(\ref{eq:projectedheisenberg}). 

Applying this to the position and Bloch momentum operators we obtain the exact operator identities
\begin{align}
    \dtRB
    &=
    \frac{\hat{\bm \pi}_{\mathcal B}}m
    +
    \ri
    \left(\R \QB \Hpert - \Hpert \QB \R\right)_{\mathcal B}
    \label{eq:constrained_dynamics_r}
    ,\\
    \dtCrMB
    &=
    q\bm E_{\mathcal B}(\R) + \frac{q}{2m}\{  [\partial_{\R}, A_j(\R)],\hat\pi_j\}_{\mathcal B}
    .
    \label{eq:constrained_dynamics_k2}
\end{align}
Here and below, an Einstein summation convention over repeated coordinate indices (in this case $j$) is used. Neither the canonical momentum $\bm{p}$ nor the crystal momentum $\bm{k}$ is gauge-invariant under a transformation of the electromagnetic potentials. In contrast, the kinematic momentum $\hat{\bm{\pi}}$ is gauge-invariant, but even in the absence of external perturbation it is not conserved due to the crystal lattice potential. Hence, a more suitable operator for studying the dynamics is the gauge-invariant crystal momentum $\hat{\bm \kappa}\equiv\hat{\bm k}-q\bm{A}(\hat{\bm r})$, which obeys the equation of motion 
\begin{align}
\dot{\hat{\bm \kappa}} = \frac q2\left(\hat{\bm\pi}\times\bm B - \bm B\times\hat{\bm\pi} + 2\bm E\right).
\end{align}  
In addition to the electric field term, it explicitly contains a Lorentz force, in analogy to the Lorentz force term appearing in ${\rm d}{\hat{\bm\pi}}/{\rm dt}$ in the absence of a lattice potential. When projected, this force equation holds approximately to first order in external fields strengths.

\Cref{eq:constrained_dynamics_r} shows that there are generically additional contributions to the band-projected velocity when the perturbation is present. They are given by
\begin{align}
    \dtRinterB
    &\equiv
    \ri \left(\R \QB \Hpert - \Hpert \QB \R\right)_{\mathcal B}
    \label{eq:inter_band_velocity}
    \\&=
    \ri
    \left[\R -\RB, \Hpert\right]_{\mathcal B}
    .
    \label{eq:inter_band_velocity_alt}
\end{align}
The corresponding term in the time derivative of the crystal momentum vanishes because $[\CrM, \PB]=0$.
Since it contains $\QB$, \cref{eq:inter_band_velocity} shows explicitly that the additional contribution is due to virtual scattering processes between the subspace of low-energy bands $\mathcal B$ and its complement.
It is therefore appropriate to call it an inter-band velocity and, as we demonstrate below, the terms that capture multi-band effects are exactly the geometric 
corrections to the electron dynamics, which include in particular the Berry curvature related anomalous velocity.  
We would like to emphasize that the band projection is crucial to find anomalous corrections: if $\mathcal B=\mathcal H$, the complement projector $\QB$ vanishes, leading to $\dtRinterB=0$.
Furthermore, the equivalent expression~\eqref{eq:inter_band_velocity_alt} demonstrates that the inter-band velocity quantifies the error of the crudest semiclassical approximation $\dtRB \approx \hat{\bm\pi}_{\mathcal B}/m$.

When projected to the subspace $\mathcal B$, many of the usual commutation relations are significantly modified. So before discussing in detail the effect of $\Hpert$, let us briefly comment on the position, canonical momentum $\hat {\bm{p}}$, kinetic momentum $\hat{\bm{\pi}}$ and gauge-invariant crystal momentum $\hat{\bm{\kappa}}$ operators. 

The position operator can be decomposed as $\hat r_j = \ri\smash{ \partial_{\hat k_j}} + \smash{\BC{j}}$ where $\smash{\BC{j}}$ is the Berry connection in operator form, which has the matrix elements $\bra{\psi_{m\bm k}}\smash{\BC{j}}\ket{\psi_{n\bm k}}=\ri \bra{u_{m\bm k}}\partial_{k_j}\ket{u_{n\bm k}} $. Although this expression was already obtained in Ref.~\cite{Blount1962FormalismsOB}, we shall provide a derivation using a moment generating operator formalism in Sec.~\ref{SM:matrix_elements_of_the_geometric_operators} of the Supplemental Material. In this expression, the operator-valued derivative $\smash{\ri\partial_{\hat k_j}} = \sum_{n\bm k}\ket{\psi_{n\bm k}}\ri\partial_{k_j} \bra{\psi_{n\bm k}}$ can be understood as the lattice position operator, while the Berry connection operator is the ``intra-cell'' position within the primitive cell~\cite{Balian1989}. It can be shown that $[\hat{r}_i,\hat{r}_j]=0$, whilst the commutator of the projected position operators no longer vanishes but produces the operator form of the Berry curvature.  Introducing the covariant derivative as $  \hat{\nabla}_j=-\ri\hat{r}_{j\mathcal{B}}=\smash{\projs{\partial_{\hat{k}_j}-\ri\hat{A}_j}}$, it is easy to derive that for an arbitrary operator with matrix elements $O_{nm}(\bm{k})$ (see Sec.~\ref{SM:matrix_elements_of_the_geometric_operators} in the Supplemental Material)  
 \begin{align}
    [\hat{\nabla}_j,\hat{O}_{\mathcal{B}}]
    %1ii
     &=
    %1ii
    \sum_{n,m\in\NB}\sum_{\bm k}
    \ket{\psi_{n\bm k}} \bra{\psi_{m\bm k}} \notag
    \\&
   \times \left(
        \partial_{k_j}O_{nm}(\bm{k})-i[\BCe{j}_{\mathcal{B}}(\bm k),O_{\mathcal{B}}(\bm k)]_{nm}
    \right)
    ,
\end{align}
where the band summations are restricted to the projected bands. When applied to the position operator itself, one recovers the usual non-Abelian Berry curvature as matrix elements. Hence we have the operator identity $[\hat{\nabla}_j,\hat{\nabla}_k]=-\ri\hat{\Omega}_{jk}$. 

When studying semiclassical equations of motion usually only the term involving the group velocity of modes is retained \cite{Ashcroft76}, which is valid if the inter-band contributions can be neglected and therefore in particular in single-band approximations. In fact, if we define the velocity operator as $\hat{\bm v}\equiv\ri[\Hbloch,\hat{r}]$, then  it can be shown that in the Bloch basis~\cite{Blount1962FormalismsOB,velocity} 
\begin{align}
    \hat{\bm v} = \sum_{nm\bm k}\ket{\psi_{n\bm k}}\bra{\psi_{m\bm k}}\left[\delta_{nm}\partial_{\bm k}\E_{n\bm k} + \ri\Delta_{nm}(\bm k)\BCebf_{nm}(\bm k)\right],
    \label{eq:velocity operator}
\end{align}
where $\Delta_{nm}(\bm k) = \E_{n\bm k}-\E_{m\bm k}$ is the direct band gap at crystal momentum $\bm k$.
The Berry connection is defined as $\BCe{nm,i}(\bm k)=\ri\braket{u_{n\bm k} | \partial_{k_i} u_{m\bm k}}$, with the cell-periodic part of the Bloch functions given by $\ket{\psi_{n\bm k}} = \exp(\ri\bm k\cdot\R)\ket{u_{n\bm k}}$. From \cref{eq:velocity operator}, it is clear that neglecting inter-band contributions corresponds to retaining only the group velocity $\partial_{\bm k}\E_{n\bm k}$. With the Bloch Hamiltonian $\Hbloch$, the canonical momentum is related to the velocity operator as $\CanM=m\hat{\bm v}$. Note this relation between the canonical momentum $\hat{\bm{p}}$ and the velocity operator $\hat{\bm{v}}$ is valid only for the specific Bloch Hamiltonian $\Hbloch$ we used because the kinetic term is quadratic. If we considered, say, a relativistic Dirac Hamiltonian in a periodic potential the expression of the velocity operator would remain valid but it can no longer be used for the canonical momentum operator. We shall give another expression of the canonical momentum below.

The position and crystal momentum operators are canonically conjugate in the sense that $[\R,\CanM] =[\R,\CrM]  = \ri$. Note that while $\CrM$ commutes with $V(\R)$ because the latter is invariant under lattice translation, $\CanM$ does not \cite{Balian1989}. Upon projection, $\R$ and $\smash{\CrM}$ remain a conjugate pair in the restricted band subspace, $[\RB,\smash{\CrM_{\mathcal{B}}}]=\ri\mathbb{1}_{\mathcal{B}}$, but $\R$ and $\hat{\bm p}$ lose this property. In fact, in analogy to the decomposition of the position operator into an inter-cell and an intra-cell part, we can express the momentum operator as $\hat{\bm p}=\hat{\bm k}-\bm  {\mathcal{\hat{A}}}^{\bm{r}}_{ \mathcal{B}} $, where the crystal momentum is the generator of inter-cell translation while the object $\bm  {\mathcal{\hat{A}}}^{\bm{r}}_{ \mathcal{B}} $, that we will call real-space Berry connection generates intra-cell translations. The operator $\bm{\mathcal{\hat{A}}}^{\bm{r}}_{ \mathcal{B}} $ has the following matrix elements in the Bloch basis
\begin{align}
    &\mathcal{A}^{\bm{r}}_{nm}(\bm{q})
   \equiv 
    \ i\frac{(2\pi)^d}{V_{uc}} \int_{\rm cell} d\bm{r}\   u^*_{n\bm{q}}(\bm r)\ \partial_{\bm{r}}u_{m\bm{q}}(\bm r)  ,
\end{align}  
or by an abuse of notation, $\mathcal{A}^{\bm{r}}_{nm}(\bm{q})=i\bra{u_{n\bm{q}}}\partial_{\bm r}\ket{u_{m\bm{q}}}$, thus justifying the name. With this we have
\begin{align}
    [\hat{  r}_{ i\mathcal{B}},\hat{  p}_{ j\mathcal{B}}]
    &=\
    \ri\mathbb{1}_{\mathcal{B}}- \ri[\hat{\nabla}_i,\hat{ {\mathcal A}}^{\bm{r}}_{j\mathcal{B}}] \label{eq:modccm}
    .
\end{align} 
where the second term spoils the canonical commutation relation, with the correction given by the real space Berry connection. 
We would like to point out that $\mathcal{A}^{\bm{r}}_{nm}(\bm{q})$ is not the same real-space Berry connection as in Refs.~\cite{changniu,tu,SHINDOU2005399}, which is related to a position-dependent basis in a self-consistent wave-packet construction. We see from the matrix elements that $\mathcal{A}^{\bm{r}}_{nm}(\bm{q})$   characterizes the spatial modulation of the Bloch states, and vanishes in the plane-wave limit. This is consistent with the fact that the crystal momentum reduces to the canonical momentum when the lattice potential is a constant, and the system describes a free particle whose Bloch states are just plane waves. We note that only the band diagonal part of the real-space Berry connection is gauge-invariant under a Bloch gauge transformation.  

{\it Inhomogeneous electric field and geometry.}
We now show how the inter-band velocity is related to quantum geometric quantities by expressing the contributions in terms of the projected quantities. 
First, we consider the case where the external perturbation is due to an electrostatic potential only.
In this case, $\Hpert=\Hpertphi$, and we find the formally exact operator relation 
\begin{align}
    \dot{\hat r}_{j\perp}^{\phi}
    &=
    -\ri
    \left[
        \hat{r}_{j\mathcal{B}}
        ,
        q\phi_{\mathcal B}(\R)
    \right]
    =
   -\frac{q}{2}
   \{\hat\Omega_{jk},
    E_{k\mathcal B}(\R)\}
    +
    \hat R_{j \perp}^\phi
    ,
    \label{eq:inter_band_velocity_phi}
\end{align}
where the first term is exactly the operator analog of the anomalous velocity and the second term $\smash{\hat R_{i \perp}^\phi}$ represents further corrections due to the spatial dependence of the electric field. We took into account that in general $\phi_{\mathcal B}(\R)\neq \phi(\RB)$ except for a uniform electric field. In fact, even if one makes the approximation $\phi_{\mathcal B}(\R)\approx \phi(\RB)$ and symmetrizes the position operators to maintain hermiticity, there would still be additional contributions besides the anomalous velocity.

In order to understand this result and simplify the general formula for $\smash{\hat r_{j\perp}^\phi}$, we calculate it order by order in a gradient expansion of the external fields.
Up to the second order in the inhomogeneity, we have
$\smash{\phi(\R) =-E^{(0)}_{ k}\hat{r}_k-E^{(1)}_{kj }\hat{r}_k\hat{r}_j/2}$, where the coefficients are symmetric in all indices and the corresponding electric field is
$\smash{E_k(\R)=E^{(0)}_{k}+ E^{(1)}_{kj} \hat{r}_j}$.
As we show in the Supplemental Material, besides the Berry curvature term there is an additional contribution to the inter-band velocity, 
\begin{align}
    \hat{R}_{j\perp}^\phi
    =
    -\frac{q}{2}E_{kl}^{(1)}
    [
            \hat{\nabla}_j
            ,
            \hat{\mathcal{G}}^{(2)}_{kl}
        ].
\end{align}
This result corresponds exactly to the operator form of the equations of motion obtained by an electron wave-packet approach in Refs.~\cite{Lapa2019,Kozii2021}, here generalized to the multi-band case. We defined a quantum metric operator by $\hat{\mathcal{G}}_{jk}=\frac{1}{2}(\hat{\mathcal Q}_{jk}^{(2)}+\hat{\mathcal Q}_{kj}^{(2)})$, where $\hat{\mathcal Q}_{jk}^{(2)}$ is the quantum geometric tensor operator, defined in Table~\ref{tab:geoquant}. We see from the expression that this geometric contribution captures the virtual scattering processes to the complement Hilbert space.

To justify our definition of operators corresponding to geometric quantities, we shall compute their matrix elements and show that they agree with the usual definitions. Here we outline a systematic way of doing such calculations, explained in more detail in the Supplemental Material. Any algebraic function of the position operator can be expressed by using the moment generating operator as
\begin{align}
    f(\hat{\bm{r}}) 
    =
    \lim_{\bm a \rightarrow \bm 0} f(\ri\partial_{\bm{a}})   \, \re^{-\ri\bm{a\cdot\hat{r}}} \,  
    ,
\end{align}
and in the Supplemental Material we show that
\begin{align}
    \re^{-\ri\bm{a\cdot\hat{r}}}
    =
    \sum_{\bm q}
    \re^{\ri\bm{q\cdot\hat{r}}}
    \hat{P}_{\bm q}
    \re^{\bm{a} \cdot \partial_{\bm{q}}}
    \hat{P}_{\bm{q}}
    \re^{-\ri \bm{q}\cdot \hat{\bm r}},
\end{align}
where we introduced the local projection operators
$
    \hat{P}_{\bm{q}}
     =
    \sum_{n} \ket{u_{n\bm{q}}} \bra{u_{n\bm{q}}}
    . 
$
Let $\hat{P}_{\bm{q}\B}$ and $\hat{Q}_{\bm{q}\B}$ denote the band projections of $\hat{P}_{\bm{q}}$ onto $\B$ and its complement, respectively. We then obtain a general result 
\begin{align}
    &   \proj{
    [\hat{r}_j, \hat{P}_{\mathcal{B}}]
    \hat{Q}_{\mathcal{B}}
    [  \hat r^{(n)}_{\bm j_n} ,\hat{P}_{\mathcal{B}}]
    }
    =
    \proj{
    \hat{r}_j \hat{Q}_{\mathcal{B}}  \hat r^{(n)}_{\bm j_n}
    }
    \\
    &
    =
    \ri^{n+1}\sum_{\bm{q}}
    e^{ \ri\bm{q}\cdot\hat{\bm{r}}} 
    \hat{P}_{\bm{q}\mathcal{B}}
    \bigg(
    \partial_{q_j}\hat{Q}_{\bm{q}\mathcal{B}}
    \bigg)
    \hat{Q}_{\bm{q}\mathcal{B}}
    \bigg(
    \partial_{q_{\bm j_n}} \hat{P}_{\bm{q}\mathcal{B}}
    e^{-\ri\bm{q}\cdot\hat{\bm{r}}} 
    \bigg) \notag
    ,
\end{align}
where we used the short-hand notation $ \hat r^{(n)}_{\bm j_n}= \hat{ r}_{ j_1} \hat{ r}_{ j_2}...\hat{ r}_{ j_n}$, and $\partial_{q_{\bm j_n}}=\partial_{q_{  j_1}}\dots\partial_{q_{ j_n}}$. In particular, for $n=1$ we have
 \begin{align}
      \hat{\mathcal Q}_{jk}^{(2)} &=  \ri ^2\proj{\left[\hat r_j, \PB\right] \QB \left[\hat r_k, \PB\right]}\notag \\
      &=\sum_{n,m\in\mathcal N_{\mathcal B}}{\mathcal Q}_{jk,nm}(\bm q)\ket{\psi_{n\bm{q}}}\bra{\psi_{m\bm{q}}},
 \end{align}
 where the matrix elements read, explicitly 
 \begin{align}
     {\mathcal Q}_{jk,nm}(\bm k)=\bra{u_{n\bm{k}}}(\partial_{k_{j}} \hat{P}_{\bm{k}\mathcal{B}}  )   (\partial_{k_{k}}\hat{P}_{\bm{k}\mathcal{B}} )\ket{u_{m\bm{k}}}, 
 \end{align}
with $k$-derivatives acting only on the local projection operator next to them. 
This is the multi-band version of the quantum geometric tensor \cite{Kozii2021}.

\begin{table}
\renewcommand\arraystretch{1.6} 
\centering
\begin{tabular}{|c|c|}
\hline
\textbf{Name} &     \textbf{Definition}   
\\
\hline
Covariant derivative        & $\hat{\nabla}_j     \equiv   -\ri\hat{r}_{j\B}$     
\\
\hline
Quantum geometric tensor         & $\hat{\mathcal Q}_{jk}^{(2)}  =  -\projs{[\hat r_j, \PB] \QB [\hat r_k, \PB]} $ 
\\      
\hline
Mixed geometric tensor 1        & $ \hat{\mathcal L}_{jk}%
    \equiv
    -\projs{
    [\hat{r}_{j  },\hat{P}_{\mathcal{B}}]
    [\hat{p}_{k  },\hat{P}_{\mathcal{B}}]
    } $   
\\ 
\hline
Mixed geometric tensor 2         & $ \hat{\mathcal{L}}_{jkl}^{(2)}\equiv \projs{[\hat{r}_k,[\hat{r}_j,\PB]]\QB[\hat{p}_l,\PB]}$    
\\ \hline
\end{tabular}
\caption{A table of the operator forms of geometric quantities.}
\label{tab:geoquant}
\end{table}

{\it Magnetic field.} After having demonstrated that our method reproduces and generalizes previous results for electric fields, we will apply the formalism now to the hitherto unknown case of an inhomogeneous magnetic field. Compared to the electric field case, a complication arises because the projected canonical momentum and position operators no longer satisfy a canonical commutation relation. The inter-band velocity operator reads
\begin{align}
    \ri\dot{\hat r}_{j\perp}^{\bm{A}} =&
     -\frac{q}{2m} \bigg( -
     \Big\{ [\hat {r}_{j\mathcal{B}}  ,  
  \hat{ {\mathcal A}}^{\bm{r}}_{k\mathcal{B}}
    ],  {A}_{k\mathcal{B}}(\bm {\hat{r}}) \Big\}   \label{eq:magfield_1}
     \\& +
      \Big\{ [\hat {r}_{j\mathcal{B}}  ,  
     {A}_{k\mathcal{B}}(\bm {\hat{r}}) 
    ],  \hat{\pi}_{k  \mathcal{B}} \Big\}   \label{eq:magfield_2}
   \\& +
   \Big[
   \hat {r}_{j\mathcal{B}}  ,  
   \proj{
     {A}_{k}(\bm {\hat{r}}) \QB  \hat p_{k  }
    +   \hat p_{k  } \QB {A}_{k }(\bm {\hat{r}})  } \Big] \bigg)
         \label{eq:magfield_3}
     \\&+\frac{q^2}{2m}
    %
    % \left(
    %
      \left[\hat{r}_{j\B}, \proj{A_{k}(\R)\QB A_{k}(\R)}\right]  
    %
    % \right) 
    .
    \label{eq:magfield_4}
\end{align}
Let us briefly comment on the origin of the terms. The first term emerges because the projected position and momentum operators are no longer a canonical pair but now satisfy \cref{eq:modccm}. \Cref{eq:magfield_2,eq:magfield_4} originate from the fact that the projected position operators no longer commute with each other. These terms can be evaluated order-by-order as we have presented for the electric field. Most interestingly, \cref{eq:magfield_3} contains new mixed geometric quantities that involve spatial gradients of the Bloch states, as we explain below.

As for the electric field case, we perform a gradient expansion of the magnetic vector potential. To the first order, $\smash{A_j(\hat{\bm{r}})= A^{(1)}_{jk}\hat{r}_k}$, the inter-band velocity can be evaluated to
\begin{align}
     \dot{\hat{r}}^{(1)}_{l\perp }  
&=
    \frac{q^2}{2m}A^{(1)}_{ka}A^{(1)}_{kb}[\hat{\nabla}_l,\hat{\mathcal G}^{(2)}_{ab}] \\
&+
    \frac{\ri q}{m} A_{jk}^{(1)}\bigg(\{ \hat{\nabla}_{k } , [\hat{\nabla}_l,\hat{ {\mathcal A}}^{\bm{r}}_{j\mathcal{B}}] \} +
     \{ \hat{\Omega}_{lk  },\hat{\pi}_{j \mathcal{B}}\}     -
     [\hat{\nabla}_{l },   \hat{\mathcal L}_{jk}] 
    \bigg) \notag.
\end{align}
The explicit expression of the next order expansion of the magnetic potential is given in the Supplemental Material. The second term in the parenthesis is the Berry curvature term appearing in standard semiclassical equation of motion \cite{changniu}. The other terms are new contributions. In particular, we introduced a mixed geometric quantity $\hat{\mathcal L}_{jk}$ which is defined, together with the other central operator-valued geometric quantities in Table~\ref{tab:geoquant}. 
Its matrix elements are evaluated to be $
     {\mathcal L}_{jk,nm}(\bm{q})
    =
    \ri \bra{u_{n\bm{q}}}
     \smash{(\partial_{q_j}
 \hat{Q}_{\bm{q}\mathcal{B}} )   
 \hat{Q}_{\bm{q}\mathcal{B}} 
 \hat{\mathcal A}^{\bm{r}}_{k\bm{q}}}
    \ket{u_{m\bm{q}}} 
$.
This is a new geometric quantity that depends on the spatial gradient of the Bloch states, embodied in the real-space Berry connection $\hat{\mathcal A}^{\bm{r}}_{j\bm{q}}\equiv e^{-\ri\bm{q}\cdot\hat{\bm r}}\hat{\mathcal A}^{\bm{r}}_{j }e^{\ri\bm{q}\cdot\hat{\bm r}}$, in addition to the momentum gradient. Moreover, in the Supplemental Material, we derive the geometric corrections to the second order in the gradient expansion, which leads to an additional term in the inter-band velocity determined by mixed geometric tensor $\hat{\mathcal L}^{(2)}_{jkl}$ shown in Table~\ref{tab:geoquant}.

%===========================================================================
%===========================================================================
{\it Semiclassical equations of motion.} So far, we have provided results at the level of operators. Next, we apply these to the semiclassical dynamics of a wavepacket constructed from states in the projected set of bands but otherwise of arbitrary shape: $\ket{W}=\sum_{n\in\NB}\int_{\rm BZ} d{\bm q} c_{n\bm{q}} \ket{\psi_{n\bm{q}}}$. The normalization is given by $\Braket{W|W}=\sum_{n}\int d{\bm q} \lvert c_{n\bm{q}}\rvert^2=1$ and we assume the state evolves with the effective Hamiltonian $\hat{H}_{\mathcal{B}}$. 

It is easy to show that the time derivative of an observable $O(t)\equiv  \bra{\psi }\hat{O}(t)\ket{\psi }$ in the Heisenberg picture can be equivalently given in the Schrodinger picture as
$\ri dO(t)/dt=\bra{\psi(t)}[\hat{O} ,\hat{H}]\ket{\psi(t)}$~\cite{Sakurai2020}. 
Then, the equations of motion of the center of mass and the gauge-invariant crystal momentum are simply the projected dynamics of the operators evaluated on the wave-packet, which read
\begin{align}
    \dot{\bm R}&=-\ri\bra{W(t)}[\hat{\bm{r}}_\mathcal{B},\hat{H}_\mathcal{B}]\ket{W(t)} %
    , \notag \\
    \dot{\bm K}&=-\ri\bra{W(t)}[\hat{\bm{\kappa}}_\mathcal{B},\hat{H}_\mathcal{B}]\ket{W(t)}, 
\end{align}

where the evolved state is parametrized as 
\begin{align}
    \ket{W(t)} =\sum_{n\in\NB}\int_{\rm BZ} d\bm q\ c_{n\bm{q}}(t) \ket{\psi_{n\bm{q}}} .
\end{align}
Now we consider the motion of the wave-packet under the influence of a magnetic field. Expanding the magnetic vector potential as $\smash{A_j(\hat{\bm{r}})= A^{(1)}_{jk}\hat{r}_k+\frac{1}{2}A^{(2)}_{jkl}\hat{r}_k\hat{r}_l}$, we focus on the paramagnetic contribution. We expand the inter-band velocity similarly order by order as $\smash{\dot{r}_{j \perp}^{\bm A}=\dot{r}^{(1)}_{j  \perp}+\dot{r}^{(2)}_{j \perp}+\ldots}$, each term corresponding to the respective term of the expansion of $A_j$. For simplicity, here we consider the single band case so that $\PB=\sum_{\bm{k}}\ket{\psi_{n\bm{k}}}\bra{\psi_{n\bm{k}}}$ for a given $n$. This applies to the case of a metallic system in which the Fermi level lies within a non-degenerate band, and the contributions due to other bands are suppressed by band gaps. The leading order contributions then read 
\begin{align}
&
    \dot{ r}^{(1)}_{l\perp} 
=
  \ri A_{jk}^{(1)} \frac{q}{m} \label{eq:wavepacket velocity}\\
& \times
  \Big\langle\!\!\Big\langle
     R_k(\bm{q},t)  \pdv{{ {\mathcal A}}^{\bm{r}}_{j\mathcal{B}}(\bm{q})}{q_l}  (\bm q)
      +{\Omega}_{lk  }(\bm{q})  {p}_{j  }(\bm{q}) 
    -
      \pdv{{\mathcal L}_{jk}(\bm{q})}{q_l}   
    \Big\rangle\!\!\Big\rangle, \notag
\end{align} 

where the double angle bracket refers to the weighted average $ \langle\!\langle ...\rangle\!\rangle\equiv\int_{\rm BZ} d\bm{ q} {c}_{\bm{q}}^\dagger(t)...{c}_{\bm{q}}(t)$. In the single-band approximation, we may use $p_j(\bm{q})/m= \partial{\epsilon(\bm{q})}/\partial{q_j} $, which is just the group velocity for the lattice Hamiltonian.  As we shall demonstrate below, the last term would be a Fermi surface contribution to the current. In the above $ R_{j}(\bm q,t)$ is the center of mass position of the wave-packet at a given time, which evaluates to be
\begin{align}
    \bra{W}\hat{r}_j(t)\ket{W}
    &=\langle\!\langle R_{j}(\bm q,t)\rangle\!\rangle
    \\&=
     \int d\bm{q}    R_{ j}(\bm q,t) 
     \lvert c_{ \bm q}(t)\rvert^2,
\end{align}
where 
$ \smash{
    R_{ j}(\bm q,t)
    \equiv
    \ri  \partial_{q_j} \theta(\bm q,t) 
    +
    \mathcal{A}_{ j}(\bm q)},
$
and $\theta(\bm q,t)$ is the phase of the expansion coefficient $c_{\bm{q}}(t)= \lvert c_{ \bm q}(t)\rvert e^{-\ri\theta(\bm q,t)} $.

%===========================================================================
%===========================================================================
{\it Sharp momentum approximation.} Here we discuss how the current due to geometric contributions may be measured in a metallic system. Within the Boltzmann transport theory, a net current may exist if there is either a nonequilibrium distribution $f(\bm{K})$ deforming the Fermi surface, or a momentum-dependent velocity term $\dot{\bm{R}}(\bm{K} )$ given by single particle dynamics beyond the group velocity, such as the Berry curvature term such as the second term in \cref{eq:wavepacket velocity} contributing to the intrinsic anomalous Hall effect~\cite{ahe} or more generally the geometric quantities we discussed above. To isolate the effects of the latter we assume an equilibrium distribution. Suppose the Fermi surface intersects only a single nondegenerate band $n$ with Bloch functions $\ket{\psi_{\bm q}}$. At zero temperature we take $f(\bm{K},t)\approx \Theta[\epsilon_{FS}-\epsilon(\bm{K})]$, where $\epsilon_{FS}$ denotes the Fermi energy. Within the semiclassical regime, the wave-packet may be taken to be sharply peaked at a single momentum so that we take $\lvert c_{\bm{q}}\rvert^2\approx \delta(\bm{q}-\bm{K})$, where $\bm{K}=\bra{W}\hat{\bm{\kappa}}\ket{W}$ is the mean gauge-invariant crystal momentum of the  wave-packet at time $t$. Then, we have the current density
\begin{align}
    \bm{J}=\int \frac{d\bm{K}}{(2\pi)^3} \dot{\bm R}(\bm{K})f(\bm{K},t),
\end{align}
where $\dot{\bm R}(\bm{K})$ is the center of mass motion of a wave-packet with mean momentum $\bm{K}$. Here we will assume the system is inversion symmetric, so $\epsilon(\bm{k})=\epsilon(-\bm{k})$. Then, we find a contribution to the current due to the mixed geometric tensor, which is given by
\begin{align}
    J_l
    &=-\ri A_{jk}^{(1)}  \frac{q}{m}\oint_{\rm FS} \frac{dS}{|\nabla_{\bm{K}}\epsilon(\bm{K})|}\pdv{\epsilon({\bm K})}{K_l}\mathcal{L}_{jk}(\bm K),
\end{align}
where the integration is over the Fermi surface. Hence, we see this is a contribution entirely from the Fermi surface, and it is determined by the mixed geometric tensor $\mathcal{L}_{jk}$. 

%===========================================================================
%===========================================================================
{\it Summary and Conclusion.} In this work, we have studied the equations of motion for particles in Bloch bands under the influence of spatially inhomogeneous electromagnetic fields.
For this purpose, we introduced an approach based on band-projected operators and the Heisenberg equation of motion, and explicitly separated anomalous contributions from the group velocity operator.
By explicitly evaluating the operator matrix elements, we demonstrated that the anomalous contributions are related to the geometry of the underlying Bloch states.
In the case of electric fields, we find results consistent with the existing literature and generalize them to the multi-band case.
Furthermore, we find that the leading order of the anomalous velocity is a product of Berry curvature and the band-projected electric field, thereby generalizing known results to a higher order in the spatial inhomogeneity of the fields.
Lastly, we address the case of spatially inhomogeneous magnetic fields and find geometric corrections to the anomalous velocity in the leading order.
We show that these leading corrections stem from the paramagnetic term and are tied to mixed geometric operators containing spatial derivatives of Bloch projectors.

%===========================================================================
%===========================================================================
{\it Acknowledgements.} We thank Christophe De Beule for stimulating discussions at the start of this project.
TLS, CX and AH acknowledge financial support from the National Research Fund Luxembourg under Grants No.~C20/MS/14764976/TopRel, INTER/17549827/AndMTI, C22/MS/17415246/DeQuSky, and AFR/23/17951349. CX, SH and TM acknowledge
funding by the Deutsche Forschungsgemeinschaft (DFG) via the Emmy Noether Programme (Quantum Design
grant, ME4844/1, project-id 327807255), project A04 of the Collaborative Research Center SFB 1143 (project-id 247310070), and the Cluster of Excellence on Complexity
and Topology in Quantum Matter ct.qmat (EXC 2147,
project-id 390858490).

\bibliography{biblio}

\clearpage\newpage
%! TeX root = draft (submission form).tex
\onecolumngrid
\begin{center}
    \textbf{\large Supplemental Material for \\``Quantum geometry in the dynamics of band-projected operators''}
\end{center}
%%%%%%%%%% Merge with supplemental materials %%%%%%%%%%
%%%%%%%%%% Prefix a "S" to all equations, figures, tables and reset the counter %%%%%%%%%%
\setcounter{equation}{0}
\setcounter{figure}{0}
\setcounter{table}{0}

\makeatletter
\renewcommand{\theequation}{S\arabic{equation}}
\renewcommand{\thefigure}{S\arabic{figure}}
\renewcommand{\thetable}{S\Roman{table}}
\newcommand{\subf}[2]{%
  {\small\begin{tabular}[t]{@{}c@{}}
  #1\\#2
  \end{tabular}}%
}

%===========================================================================
\section{Band-projected products of position operators and quantum geometry}
\label{SM:projectedMGF}
%===========================================================================
Here, we motivate the covariant derivative from the band-projected kinematic momentum derivative, i.e., $\RB = \ri\PB \partial_{\hat {\bm p}} \PB = \ri\hat{\bm\nabla}$ and find its associated geometric objects by expressing products of band-projected position operators in terms of $\hat{\bm\nabla}$.
We can use the Bloch basis to give a precise meaning to the covariant derivative,
\begin{align}
    \hat\nabla_i = \partial_{\hat k_i,\mathcal B} - \ri\BCB{i}
    ,\quad
    \partial_{\hat k_i,\mathcal B} = \sum_{n\in\mathcal N_{\mathcal B}}\sum_{\bm k}\ket{\psi_{n\bm k}}\partial_{k_i}\bra{\psi_{n\bm k}}
    ,\quad
    \BCB{i} = \sum_{n,m\in\mathcal N_{\mathcal B}}\sum_{\bm k}\BCe{nm,i}(\bm k)\ket{\psi_{n\bm k}}\bra{\psi_{m\bm k}}
    ,
\end{align}
with the definition of the Berry connection $\BCe{nm,i}(\bm k) = \ri\bra{u_{n\bm k}}  \partial_{k_i} \ket{ u_{m\bm k}}$. The band-projected quadrupole moment reads
\begin{align}
    \left(\hat r^{(2)}_{ij}\right)_{\mathcal B}
    =
    \left(\hat r_i\hat r_j\right)_{\mathcal B}
    =
    \hat r_{i,\mathcal B}\hat r_{j,\mathcal B}
    +
    \left(\hat r_i \QB \hat r_j\right)_{\mathcal B}
    =
    \hat r_{i,\mathcal B}\hat r_{j,\mathcal B}
    -
    \left(\left[\hat r_i,\PB\right] \QB \left[\hat r_j,\PB\right] \right)_{\mathcal B}
    =
    \ri^2
    \left(
        \hat \nabla_i\hat \nabla_j
        -
        \hat{\mathcal Q}^{(2)}_{ij}
    \right)
    ,
\end{align}
which contains the quantum geometric tensor $\hat{\mathcal Q}_{ij}^{(2)}=-\projs{[\hat r_i,\PB] \QB [\hat r_j,\PB]}$. In the following, we repeatedly exploit the idempotence of $\PB$ and $\QB$, which implies that the following expressions vanish for $N\in\mathbb N$,
\begin{align}
    \PB \prod_{n=1}^{2N-1}\left[\hat O_n, \PB\right]\PB = 0
    ,\quad
    \QB \prod_{n=1}^{2N}\left[\hat O_n, \PB\right]\PB = 0
    .
\end{align}
In general, we relate the $n$th moment of the position operator to a lower order by inserting $\mathbb 1=\PB + \QB$, i.e.
\begin{align}
    \PB \hat r^{(n)}_{i_1\dots i_n}\PB
    %1i
    &=
    %1i
    \PB \hat r_{i_1}\left(\PB + \QB\right)\hat r^{(n-1)}_{i_2\dots i_n}\PB
    %1ii
    =
    %1ii
    \hat r_{i_1,\mathcal B}\hat r^{(n-1)}_{i_2\dots i_n,\mathcal B}
    +
    \PB \hat r_{i_1}\QB\hat r^{(n-1)}_{i_2\dots i_n}\PB
    \\
    %2
    &=
    %2
    \hat r_{i_1,\mathcal B}\hat r^{(n-1)}_{i_2\dots i_n,\mathcal B}
    -
    \PB
    \left[\hat r_{i_1},\PB\right]
    \QB
    \left[\hat r^{(n-1)}_{i_2\dots i_n}, \PB\right]
    \PB
    .
    \label{eq:band_projected_position_operators}
\end{align}
Note that operators associated with the Bloch geometry emerge from the second term of \cref{eq:band_projected_position_operators}. This can be brought forward by using the following recursive relation
\begin{align}
    \QB \left[\hat r^{(n)}_{i_1\dots i_n}, \PB\right] \PB
    &=
    \QB
    \left(
        \left[
            \hat r_{i_1},
            \QB
            \left[
                \hat r^{(n-1)}_{i_2\dots i_n}
                ,
                \PB
            \right]
            \PB
        \right]
        +
        \left[
            \hat r^{(n-1)}_{i_2\dots i_n}
            ,
            \PB
        \right]
        \hat r_{i_1,\mathcal B}
        +
        \left[
            \hat r_{i_1},
            \PB
        \right]
        \hat r^{(n-1)}_{i_2\dots i_n,\mathcal B}
    \right)
    \PB
    ,
    \label{eq:recursive_commutator}
\end{align}
where in the first term of \cref{eq:recursive_commutator}, we find that
\begin{align}
    \QB\left[
    \hat r_{i_1},
    \QB
    \left[
        \hat r^{(n-1)}_{i_2\dots i_n}
        ,
        \PB
    \right]
    \PB
    \right]\PB
    =
    \QB\left[
    \hat r_{i_1},
    % \QB
    \left[
        \hat r^{(n-1)}_{i_2\dots i_n}
        ,
        \PB
    \right]
    % \PB
    \right]\PB
    .
\end{align}
This motivates the definition of the $n$th geometric quantity through a recursive commutator
\begin{align}
    \hat{\mathcal Q}^{(n)}_{i_1i_2\dots i_n}
    &=
    (-\ri)^n
    \PB
    \left[\hat r_{i_1},\PB\right]
    \QB
    \left[\hat r_{i_2},\left[\dots,\left[\hat r_{i_n},\PB\right]\right]\right]
    \PB
    .
    \label{eq:nth_geometric_operator}
\end{align}
Note that $\QB\rightarrow \PB$ in \cref{eq:nth_geometric_operator} leads to a vanishing expression, which highlights that geometric operators exclusively stem from virtual transitions between the projected Hilbert space and its complement.
Using \cref{eq:band_projected_position_operators}, \cref{eq:recursive_commutator} and \cref{eq:nth_geometric_operator}, we readily evaluate the product of three band-projected position operators and find
\begin{align}
    \left(\hat r^{(3)}_{ijk}\right)_{\mathcal B}
    =
    \left(\hat r_i\hat r_j \hat r_k\right)_{\mathcal B}
    =
    \ri^3
    \left(
        \hat\nabla_i\hat\nabla_j\hat\nabla_k
        -
        \left(
            \hat{\mathcal Q}^{(2)}_{ij}\hat\nabla_k
            +
            \hat{\mathcal Q}^{(2)}_{ik}\hat\nabla_j
            +
            \hat{\mathcal Q}^{(2)}_{jk}\hat\nabla_i
            +
            \left[\hat\nabla_i, \hat{\mathcal Q}^{(2)}_{jk}\right]
            +
            \hat{\mathcal Q}^{(3)}_{ijk}
        \right)
    \right)
    ,
\end{align}
which includes the quantum geometric tensor $\hat{\mathcal Q}^{(2)}$ and the next geometric object $\hat{\mathcal Q}^{(3)}$, which is the quantum geometric connection introduced in Refs.~\cite{Ahn2020TheoryOO,Kozii2021,Avdoshkin2023}.

We proceed by index-symmetrization of the previous expressions.
Note that we can write
\begin{align}
    \hat r^{(2)}_{ij,\mathcal B}
    =
    \hat r_{i,\mathcal B}\hat r_{j,\mathcal B} + \hat{\mathcal Q}^{(2)}_{ij}
    =
    \frac12\left\{\hat r_{i,\mathcal B},\hat r_{j,\mathcal B}\right\}
    +
    \left(
        \hat{\mathcal Q}^{(2)}_{ij}
        +
        \frac12\left[\hat r_{i,\mathcal B},\hat r_{j,\mathcal B}\right]
    \right)
\end{align}
where the commutator $[\hat r_{i,\mathcal B}, \hat r_{j,\mathcal B}] = \ri^2[\hat \nabla_i, \hat \nabla_j] = \ri\hat\Omega_{ij}$ is the curvature defined in the $\mathcal B$ subspace,
\begin{align}
    \hat \Omega_{ij}
    =
    -\ri[\hat r_{i,\mathcal B}, \hat r_{j,\mathcal B}]
    =
    \ri
    \left[
        \partial_{\hat k_i,\mathcal B} - \ri\BCB{i}
        ,
        \partial_{\hat k_j,\mathcal B} - \ri\BCB{j}
    \right]
    =
    \left[
        \partial_{\hat k_i,\mathcal B}
        ,
        \BCB{j}
    \right]
    -
    \left[
        \partial_{\hat k_j,\mathcal B}
        ,
        \BCB{i}
    \right]
    +
    \ri
    \left[
        \BCB{i}
        ,
        \BCB{j}
    \right]
    .
\end{align}
The curvature is the anti-symmetric part of the quantum geometric tensor, which can be seen from
\begin{align}
    \hat{\mathcal Q}^{(2)}_{ij} - \hat{\mathcal Q}^{(2)}_{ji}
    =
    \PB
    \left[
        \left[\PB,\hat r_i\right]
        ,
        \left[\hat r_j,\PB\right]
    \right]
    \PB
    =
    -
    \left[
        \hat r_{i,\mathcal B}
        ,
        \hat r_{j,\mathcal B}
    \right]
    =
    -
    \ri\hat\Omega_{ij}
    .
\end{align}
We define the quantum metric as $\hat{\mathcal G}_{ij}=\frac12(\hat{\mathcal Q}^{(2)}_{ij}+\hat{\mathcal Q}^{(2)}_{ji})$, therefore we arrive at the following relations
\begin{align}
    \hat{\mathcal Q}^{(2)}_{ij}
    =
    \hat{\mathcal G}_{ij} + \frac12[\RBc{i},\RBc{j}] = \hat{\mathcal G}_{ij} - \frac\ri2\hat\Omega_{ij}
    ,\quad
    \left(\hat r^{(2)}_{ij}\right)_{\mathcal B}
    =
    \ri^2
    \left(
        \frac12\{\hat\nabla_i,\hat\nabla_j\}
        -
        \hat{\mathcal G}_{ij}
    \right)
    \label{eq:2nd_geometric_operators}
    .
\end{align}
Similarly, the projected product of three position operators can be cast into an explicitly symmetric form, i.e.
\begin{align}
    \left(r^{(3)}_{ijk}\right)_{\mathcal B}
    %1
    &=
    %1
    \frac16\left(\RBc{j}\RBc{i}\RBc{k} + \text{permutations of }ijk\right)
    +
    \frac12
    \left(
        \{\hat{\mathcal G}_{ij},\RBc{k}\}
        +
        \{\hat{\mathcal G}_{ik},\RBc{j}\}
        +
        \{\hat{\mathcal G}_{jk},\RBc{i}\}
    \right)
    %2
    \\
    &\quad\quad\quad\quad
    %2
    +
    \ri
    \left(
        Q^{(3)}_{ijk}
        +
        \frac\ri2
        \left(
            \left[\RBc{k}, \hat{\mathcal G}_{ij}\right]
            +
            \left[\RBc{j}, \hat{\mathcal G}_{ik}\right]
            -
            \left[\RBc{i}, \hat{\mathcal G}_{jk}\right]
        \right)
        +
        \frac16
        \left(
            \left[\RBc{k}, \hat\Omega_{ij}\right]
            +
            \left[\RBc{j}, \hat\Omega_{ik}\right]
        \right)
    \right)
    \label{eq:projected_r3_last_term}
    .
\end{align}
To bring forward the symmetric tensor in $\hat{\mathcal Q}^{(3)}$, it is useful to define the following two operators, i.e.
\begin{align}
    \hat{\mathcal Q}^{(3)}_{ijk}
    %1
    &=
    %1
    \hat \Gamma_{i,jk}
    -
    \frac\ri2
    \hat{\tilde\Gamma}_{i,jk} \label{eq:q3}
    ,
    \\  
    \hat \Gamma_{i,jk}
    %2i
    &=
    %2i
    \frac12
    \left(
        \left[\hat\nabla_k, \hat{\mathcal G}_{ij}\right]
        +
        \left[\hat\nabla_j, \hat{\mathcal G}_{ik}\right]
        -
        \left[\hat\nabla_i, \hat{\mathcal G}_{jk}\right]
    \right)  
    %2ii
    ,\
    \hat{\tilde\Gamma}_{i,jk}
    =
    %2ii
    \frac13
    \left(
        \left[\hat\nabla_k,\hat\Omega_{ij}\right]
        +
        \left[\hat\nabla_j,\hat\Omega_{ik}\right]
    \right)
    +
    2\hat{\mathcal T}_{ijk}
    \label{eq:christoffel_symbols}
    ,
\end{align}
which reduces \cref{eq:projected_r3_last_term} to the symmetric tensor $\hat{\mathcal T}_{ijk}$. 
From \cref{eq:christoffel_symbols}, we can derive the following relation:
\begin{align}
    \left[\hat\nabla_i,\hat\Omega_{jk}\right]
    =
    \hat{\tilde\Gamma}_{j,ki}-\hat{\tilde\Gamma}_{k,ji}
    \label{eq:metric_compatibility}
    .
\end{align}
If the (matrix) inverse of $\hat\Omega$ exists (in the following denoted by upper indices), we define
\begin{align}
    \hat{\tilde\Gamma}^i_{jk} = \hat\Omega^{il}\hat{\tilde\Gamma}_{l,jk}
\end{align}
as the components of the symplectic connection $\hat{\nabla}^\prime$ which satisfies $\hat{\nabla}^\prime_i\hat\Omega_{jk} = [\hat\nabla_i,\hat\Omega_{jk}] + \hat\Omega_{jl}\hat{\tilde\Gamma}^{l}_{ki} + \hat\Omega_{lk}\hat{\tilde\Gamma}^{l}_{ji} = 0$.

Similar structures emerge in the band-projected moments containing more than three position operators and can be obtained in a straightforward way by the illustrated decomposition into quantum geometric objects, followed by index symmetrization.
These results are compactly represented by a moment generating function $\hat M(\bm\lambda) = \exp(\bm\lambda\cdot\R)$, and it's band-projected version
\begin{align}
    \left(
        \hat r^{(n)}_{\bm i_n}
    \right)_{\mathcal B}
    =
    \left(
        \hat r_{i_1}
        \dots
        \hat r_{i_n}
    \right)_{\mathcal B}
    =
    \lim_{\bm\lambda \rightarrow \bm 0}
    \partial_{\lambda_{\bm i_n}}
    \hat M_{\mathcal B}(\bm\lambda)
    ,\quad
    \hat M_{\mathcal B}(\bm \lambda)
    =
    \left(\re^{\bm\lambda\cdot\R}\right)_{\mathcal B}
    =
    \exp
    \left(
        \lambda_i\RBc{i}
        +
        \sum_{n=2}^\infty\frac1{n!}\lambda_{\bm i_n}\hat{\mathcal T}^{(n)}_{\bm i_n}
    \right)
    ,
    \label{eq:projectedMGF}
\end{align}
where $\partial_{\lambda_{\bm i_n}} = \partial_{\lambda_{i_1}}\dots\partial_{\lambda_{i_n}}$, $\lambda_{\bm i_n}=\lambda_{i_1}\dots\lambda_{i_n}$ and $\hat{\mathcal T}^{(n)}$ denotes the fully symmetric rank-$n$ tensor contained in $\hat{\mathcal Q}^{(n)}$.
In the argument of the moment generating function, we use the sum convention, including the multi-index $\bm i_n = i_1\dots i_n$.
Note that with this definition, $\hat{\mathcal G}_{ij} = \hat{\mathcal T}^{(2)}_{ij}$ and $\hat{\mathcal T}_{ijk} = \hat{\mathcal T}^{(3)}_{ijk}$.
We cannot provide a complete proof at this time, but the systematic steps which have led us to this result allow us to conjecture that \cref{eq:projectedMGF} is exact.

Using the algebraic identity
\begin{align}
    \left[\hat A, \re^{\hat B}\right]
    =
    \int_0^1\rd s \re^{(1-s)\hat B}[\hat A, \hat B] \re^{s\hat B}
    ,
\end{align}
and $\lambda_j\hat M_{\mathcal B}(\bm\lambda) = \projs{[\partial_{\hat r_j},\hat M(\bm\lambda)]}$, we find 
\begin{align}
    \frac1{\ri}\left[\RBc{i},\hat M_{\mathcal B}(\bm\lambda)\right]
    \hat\Omega_{ij}
    \left(
        \left[
            \partial_{\hat r_j}
            ,
            \hat M(\bm\lambda)
        \right]
    \right)_{\mathcal B}
    +
    \lambda_j
    \int\limits_0^1\rd s
    \left[
        \hat M^{1-s}_{\mathcal B}(\bm\lambda)
        ,
        \hat\Omega_{ij}
    \right]
    \hat M^s_{\mathcal B}(\bm\lambda)
    +
    \sum_{n=2}^\infty\frac1{n!}\lambda_{\bm i_n}
    \int\limits_0^1\rd s
    \hat M^{1-s}_{\mathcal B}(\bm\lambda)
    \left[
        \hat\nabla_i,
        \hat{\mathcal T}^{(n)}_{\bm i_n}
    \right]
    \hat M^s_{\mathcal B}(\bm\lambda)
    ,
\end{align}
where we use the short-hand notation
\begin{align}
    \hat M^s_{\mathcal B}(\bm\lambda) = \exp
    \left[
        s
        \left(
            \lambda_i\RBc{i}
            +
            \sum_{n=2}^\infty\frac1{n!}\lambda_{\bm i_n}\hat{\mathcal T}^{(n)}_{\bm i_n}
        \right)
    \right]
    .
\end{align}
For any function that admits a Taylor series expansion of the form
\begin{align}
    f(\R) = \sum_{n=0}^\infty \frac1{n!} f^{(n)}_{\bm i_n} \hat r^{(n)}_{\bm i_n}
    ,\quad
    f^{(n)}_{\bm i_n} = \partial_{x_{i_1}}\dots\partial_{x_{i_n}} f(\bm x)\big|_{\bm x = \bm 0}
    ,
\end{align}
we can directly relate the band-projected function to the moment generating function as follows:
\begin{align}
    f_{\mathcal B}(\R)
    =
    \sum_{n=0}^\infty \frac1{n!} f^{(n)}_{\bm i_n} \left(\hat r^{(n)}_{\bm i_n} \right)_{\mathcal B}
    =
    \lim\limits_{\bm\lambda\rightarrow\bm0} \sum_{n=0}^\infty \frac1{n!} f^{(n)}_{\bm i_n}\partial_{\lambda_{\bm i_n}}\hat M_{\mathcal B}(\bm\lambda)
    ,
\end{align}
and find an exact relation for its commutator with the band-projected position operator
\begin{align}
    \frac1\ri
    \left[
        \RBc{i}
        ,
        f_{\mathcal B}(\R)
    \right]
    &=
    \hat\Omega_{ij}
    \left(
        \left[
            \partial_{\hat r_j}
            ,
            f(\R)
        \right]
    \right)_{\mathcal B}
    +
    \hat R_{i, f, \mathcal B}
\end{align}
with a function-dependent remainder
\begin{align}
    \hat R_{i, f, \mathcal B}
    &=
    \lim\limits_{\bm\lambda\rightarrow\bm0}
    \sum_{n=0}^\infty \frac1{n!}
    f^{(n)}_{\bm i_n}\partial_{\lambda_{\bm i_n}}
    \left(
        \int\limits_0^1\rd s
        \left[
            \hat M^{1-s}_{\mathcal B}(\bm\lambda)
            ,
            \hat\Omega_{ij}
        \right]
        \lambda_j
        \hat M^s_{\mathcal B}(\bm\lambda)
        +
        \sum_{k=2}^\infty\frac1{k!}\lambda_{\bm j_k}
        \int\limits_0^1\rd s
        \hat M^{1-s}_{\mathcal B}(\bm\lambda)
        \left[
            \hat\nabla_i,
            \hat{\mathcal T}^{(k)}_{\bm j_k}
        \right]
        \hat M^s_{\mathcal B}(\bm\lambda)
    \right)
    ,
\end{align}
where we again use the sum convention for the multi-indices $\bm i_n$, $\bm j_n$ as well as the index $j$.

%===================================================================================================================

%===========================================================================
\section{Matrix elements of the geometric operators}\label{SM:matrix_elements_of_the_geometric_operators}
%===========================================================================
%===========================================================================
\subsection{Local projection operators and moment generating function}
%===========================================================================
To make contact with the conventional expressions for the geometric quantities, we expand the geometric operators in the Bloch basis.
We introduce the local projection operators
\begin{align}
    \hat{P}_{\bm{q}}
    &=
    \sum_{n} \ket{u_{n\bm{q}}} \bra{u_{n\bm{q}}}
    ,\quad
    \hat{P}_{\bm{q}\mathcal{B}}
    =
    \sum_{n\in \mathcal N_{\mathcal B}} \ket{u_{n\bm{q}}} \bra{u_{n\bm{q}}}
    ,\quad
    \hat{Q}_{\bm{q}\mathcal{B}}
    =
    \sum_{n\notin \mathcal N_{\mathcal B}}\ket{u_{n\bm{q}}}\bra{u_{n\bm{q}}}
    =
    \hat{P}_{\bm{q}} - \hat{P}_{\bm{q}\mathcal{B}}
\end{align}
where $\hat{P}_{\bm{q}\mathcal{B}}$ and $\hat{Q}_{\bm{q}\mathcal{B}}$ are idempotent and orthogonal, i.e. $\hat P_{\bm q\mathcal B}^2=\hat P_{\bm q\mathcal B}$, $\hat Q_{\bm q\mathcal B}^2=\hat Q_{\bm q\mathcal B}$, $\hat P_{\bm q\mathcal B}\hat Q_{\bm q\mathcal B} = \hat Q_{\bm q\mathcal B}\hat P_{\bm q\mathcal B} = 0$.
We have also that
\begin{align}
    \mathbb 1
    =
    \sum_{\bm q} \re^{\ri\bm{q\cdot\hat{r}}} \hat{P}_{\bm{q}} \re^{-\ri\bm{q\cdot\hat{r}}}
    ,\quad
    \hat{P}_{\mathcal{B}}
    =
    \sum_{\bm q} \re^{\ri\bm{q\cdot\hat{r}}} \hat{P}_{\bm{q}\mathcal{B}} \re^{-\ri\bm{q\cdot\hat{r}}}
    .
    % \label{resof1}
\end{align}
The local expressions of general projected and geometric quantities can be computed by using
\begin{align}
    O(\hat{\bm{r}})_{\mathcal{B}}
    =
    \lim_{\bm a \rightarrow \bm 0} O(\ri\partial_{\bm{a}}) \PB \, \re^{-\ri\bm{a\cdot\hat{r}}} \, \PB
    ,
\end{align}
and the identities
\begin{align}
    \PB \, \re^{-\ri\bm{a\cdot\hat{r}}} \, \PB
    %1
    &=
    %1
    \sum_{\bm{q}}
        \re^{\ri\bm q\cdot\hat{\bm r}} \, \hat{P}_{\bm{q}\mathcal{B}} \, \re^{ \bm a \cdot \partial_{\bm q}} \,
        \hat{P}_{\bm q\mathcal{B}} \, \re^{-\ri\bm{q\cdot\hat{r}}}
    ,
    \label{eq:expandedmom}
    %2
    \\
    \QB \, \re^{-\ri\bm a\cdot\hat{\bm r}} \, \PB
    &=
    %2
    \sum_{\bm{q}}
        \re^{\ri\bm{q\cdot\hat{r}}} \, \hat{P}_{\bm{q}\mathcal{B}} \, \re^{\bm{a}\cdot\partial_{\bm{q}}} \,
        \hat{Q}_{\bm{q}\mathcal{B}}  \, \re^{-\ri\bm{q\cdot\hat{r}}}
    \label{eq:expandedmom2}
    ,
\end{align}
where the exponential of derivatives acts on everything to the right.

The proof of the above formulas can be done in the position basis and by using the Poisson summation formula. For the matrix element in Bloch basis we have 
\begin{align}
    \bra{\psi_{n\bm{q}}} \re^{-\ri\bm{a\cdot\hat{r}}} \ket{\psi_{m\bm{k}}}
    &=
    \bra {u_{n\bm{q}} } re^{\ri\left(\bm k - \left(\bm a + \bm q\right)\right)\cdot\hat{\bm r}} \ket{u_{m\bm{k}}}
    \notag \\
&=
    \braket{u_{n \bm q} | u_{m \bm k}}\delta(\bm a + \bm q - \bm k)
    \equiv
    \braket{u_{n\bm{q}}| u_{m\bm{k}}}\re^{\bm{a} \cdot \partial_{\bm{q}}} \delta(\bm{q} - \bm{k}),  
\end{align}
where in the last step we used the convention that the inner product between  the cell-periodic part of the   Bloch states are  given as an integration over cells. Note that, compared to the convention in \cref{eq:projectedMGF}, we choose $\bm a = \ri \bm \lambda$ here, which brings forward the role of the auxiliary field $\bm a$ as a shift of the crystal momentum $\bm q$ in the Bloch basis. In performing the Poisson summation we also used that the Bloch momenta are restricted to the first Brillouin zone.
Using the matrix element, it is then straightforward to show
\begin{align}
    \re^{-\ri\bm{a\cdot\hat{r}}}
    =
    \sum_{\bm q}
    \re^{\ri\bm{q\cdot\hat{r}}}
    \hat{P}_{\bm q}
    \re^{\bm{a} \cdot \partial_{\bm{q}}}
    \hat{P}_{\bm{q}}
    \re^{-\ri \bm{q}\cdot \hat{\bm r}}.
    \label{eq:unprojmoment}
\end{align}
By applying band projection operators accordingly we immediately obtain \Cref{eq:expandedmom,eq:expandedmom2}.

%===========================================================================
\subsection{Position and velocity operator}
\label{sec:position_and_canonical_momentum}
%===========================================================================
We now show explicitly that the position operator can be decomposed as
\begin{align}
    \hat r_i = i\partial_{\hat k_i } +\BC{i},
\end{align}
where
\begin{align}
    \quad
    \partial_{\hat k_i } = \sum_{n }\sum_{\bm k}\ket{\psi_{n\bm k}}\partial_{k_i}\bra{\psi_{n\bm k}}
    ,\quad
    \BC{i} = \sum_{nm}\sum_{\bm k}\ket{\psi_{n\bm k}}\BCe{nm,i}(\bm k)\bra{\psi_{m\bm k}}
    .
\end{align}
Using \cref{eq:unprojmoment} we have, (derivatives act on everything to the right)
\begin{align}
    \hat{r}_i
    %1i
    &=
    %1i
    \lim_{\bm a\rightarrow \bm 0} \ri\partial_{a_i} \re^{-\ri\bm a\cdot\hat{\bm r}}
    %1ii
    =
    %1ii
    \ri\sum_{\bm k} \re^{\ri\bm k \cdot \hat{\bm r}} \hat{P}_{\bm k} \partial_{k_i} \hat{P}_{\bm k} \re^{-\ri\bm k \cdot \hat{\bm r}}
    %1iii
    =
    %1iii
    \ri \sum_{nm} \sum_{\bm k}
    \ket{\psi_{n\bm k}}
    \bra{u_{n\bm k}} \partial_{k_i}
    \ket{u_{m \bm k}}
    \bra{\psi_{m \bm k}}
    \label{eq:position_operator_matrix_elements_1}
    %2
    \\
    &=
    %2
    \sum_{nm}\sum_{\bm k}
    \ket{\psi_{n \bm k}}
    \left(
        \delta_{nm}\partial_{k_i} + A_{nm,i}(\bm k)
    \right)
    \bra{\psi_{m\bm k}}
    ,
\end{align}
where $\partial_{k_i}$ in \cref{eq:position_operator_matrix_elements_1} acts on $\hat P_{\bm k}\re^{-\ri\bm k\cdot\hat{\bm r}} = \sum_m \ket{u_{m\bm k}}\bra{\psi_{m\bm k}}$, and $\bm{A}_{nm}(\bm k) =\ri \bra{u_{n \bm k}} \partial_{\bm k} \ket{u_{m \bm k}}$ is the non-abelian Berry connection. It can be shown that when un-projected $\partial_{k_i}\hat{A}_j-\partial_{k_j}\hat{A}_i=[\hat{A}_i,\hat{A}_j]$, which is consistent with the fact that the position operators  commute, $[\hat{r}_i,\hat{r}_j]=0$. This motivates us to denote the projected position operator as a covariant derivative $\hat{r}_{i\mathcal{B}}=\ri \hat{\nabla}_i$.
\\
\\
   A general  Bloch diagonal operator is of the form $\hat{O}= \sum_{n }\sum_{\bm k}\ket{\psi_{n\bm k}}O_{nm}(\bm{k})\bra{\psi_{m\bm k}}$, and we have
\begin{align}
    \frac{1}{i}[\hat r_i, \hat{O}]
    =
    \sum_{nm}\sum_{\bm k}
    \ket{\psi_{n\bm k}}
    \left(
        \partial_{k_i}O_{nm}(\bm{k})-i[\BCe{i}(\bm k),O(\bm k)]_{nm}
    \right)
    \bra{\psi_{m\bm k}}
    ,
\end{align}
where the derivative acts on $O_{nm}(\bm k)$ only, and $[\BCe{i}(\bm k),O(\bm k)]_{nm}=\sum_{l}\BCe{i,nl}(\bm k)  O_{lm}(\bm k)-O_{nl}(\bm k)\BCe{i, lm}(\bm k)$. Note that the second term vanishes for an operator whose matrix elements are band-independent.
Similarly, for band-projected Bloch diagonal operators, we have
\begin{align}
    \frac{1}{i}[\hat{r}_{i\mathcal{B}},\hat{O}_{\mathcal{B}}]
    %1i
    &=
    %1i
    [\hat{\nabla}_i,\hat{O}_{\mathcal{B}}]
    %1ii
    =
    %1ii
    \sum_{n,m\in\NB}\sum_{\bm k}
    \ket{\psi_{n\bm k}}
    \left(
        \partial_{k_i}O_{nm}(\bm{k})-i[\BCe{i}_{\mathcal{B}}(\bm k),O_{\mathcal{B}}(\bm k)]_{nm}
    \right)
    \bra{\psi_{m\bm k}}
    %2
    \\
    &=
    %2
    \sum_{n,m\in\NB}\sum_{\bm k}
    \ket{\psi_{n\bm k}}
    [\nabla_i,O_{\mathcal{B}}(\bm k)]_{nm}
    \bra{\psi_{m\bm k}}
    ,
\end{align}
where the band summation is now restricted to the projected bands. In particular, we have
\begin{align}
    [\hat{r}_i,\hat{k}_j]
    =
    \ri
    [\partial_{\hat{k}_i},\hat{k}_j]
    =
    \ri
    \delta_{ij}
\end{align}
 With these expressions, we can also evaluate the velocity operator to be 
\begin{align}
    \hat{\bm v} 
    =
    \frac{1}{\ri}
    [\hat{\bm r}, \hat{H}_0]
    =
    \sum_{nm\bm k}
    \ket{\psi_{n\bm k}}
    \bra{\psi_{m\bm k}}
    \left(
        \delta_{nm}\partial_{\bm k}\E_{n\bm k} + \ri\Delta_{nm}(\bm k)\BCebf_{nm}(\bm k)
    \right),
\end{align}
which gives \cref{eq:velocity operator} in the main text. Note with the specific crystal Hamiltonian $\Hbloch=(\CanM^2/2m) + V(\R) $, the velocity operator is simply $\hat{\bm v}= \frac{\hat{\bm{p}}}{m}$, which for instance, for the Pauli Hamiltonian with a spin orbit coupling or a Dirac electron moving in periodic potential, this no longer holds\cite{Blount1962FormalismsOB}.

%===========================================================================
\subsection{Geometric quantities}
%===========================================================================
We would now compute the matrix elements of the operators associated with the geometric quantities, and show that they are consistent with the conventional ones seen in literature. We do this by using \cref{eq:unprojmoment}, which by applying derivatives gives 
\begin{align}
    &   \proj{
    [\hat{r}_i, \hat{P}_{\mathcal{B}}]
    \hat{Q}_{\mathcal{B}}
    [  \hat r^{(n)}_{\bm i_n} ,\hat{P}_{\mathcal{B}}]
    }
    =
    \proj{
    \hat{r}_i \hat{Q}_{\mathcal{B}}  \hat r^{(n)}_{\bm i_n}
    }
    \\
    &
    =
    \ri^{n+1}\sum_{\bm{k}}
    e^{ i\bm{k}\cdot\hat{\bm{r}}} 
    \hat{P}_{\bm{k}\mathcal{B}}
    \bigg(
    \partial_{k_i}\hat{Q}_{\bm{k}\mathcal{B}}
    \bigg)
    \hat{Q}_{\bm{k}\mathcal{B}}
    \bigg(
    \partial_{k_{i_1}}\dots\partial_{k_{i_n}}\hat{P}_{\bm{k}\mathcal{B}}
    e^{-i\bm{k}\cdot\hat{\bm{r}}} 
    \bigg)
    ,
\end{align}
where as in the previous section $ \hat r^{(n)}_{\bm i_n}= \hat{ r}_{ i_1} \hat{ r}_{ i_2}...\hat{ r}_{ i_n}$.  
%===========================================================================
\subsubsection{Quantum geometric tensor}\label{SM: quantum geometric tensor}
%===========================================================================
For $n=1$ we have
 \begin{align}
      \hat{\mathcal Q}_{ij}^{(2)} &=  \ri ^2\proj{\left[\hat r_i, \PB\right] \QB \left[\hat r_j, \PB\right]}\\
      &=\sum_{n,m\in\mathcal N_{\mathcal B}}{\mathcal Q}_{ij,nm}(\bm k)\ket{\psi_{n\bm{k}}}\bra{\psi_{m\bm{k}}},
 \end{align}
 where  the matrix element reads explicitly 
 \begin{align}
     {\mathcal Q}_{ij,nm}(\bm k)=\bra{u_{n\bm{k}}}\bigg(\partial_{k_{i}} \hat{P}_{\bm{k}\mathcal{B}}  \bigg)   \bigg(\partial_{k_{j}}\hat{P}_{\bm{k}\mathcal{B}} \bigg)\ket{u_{m\bm{k}}},\label{eq:qglocal}
 \end{align}
 with $k$-derivatives that act only on the closest local projection operators. 
This is the multi-band version of quantum geometric tensor \cite{Kozii2021}. We note that the quantum geometric tensor is also the projected cumulant $\hat{Q}_{ij}^{(2)}=\hat{P_\mathcal{B}}(\hat{r}_i-\hat{r}_{i\mathcal{B}})(\hat{r}_j-\hat{r}_{j\mathcal{B}})\hat{P_\mathcal{B}}$. \cref{eq:qglocal} is exactly the usual from of quantum geometric tensor\cite{mera} , whose real and imaginary parts correspond to the quantum metric and Berry curvature tensor respectively: $ {\mathcal Q}_{ij,nm}(\bm k)={\mathcal G}_{ij,nm}(\bm k) -\frac{\ri}{2}\Omega_{ij,nm}(\bm k)$. 

%===========================================================================
\subsubsection{ Christoffel symbols and symplectic connection}
%===========================================================================
\begin{align}
    & \hat{P}_{\mathcal{B}}[\hat{r}_i, \hat{P}_{\mathcal{B}}]\hat{Q}_{\mathcal{B}} [ \hat{r}_{j } \hat{r}_{ k},\hat{P}_{\mathcal{B}}]\hat{P}_{\mathcal{B}} \\
    &=i^{3}\sum_{\bm{k}}e^{ i\bm{k}\cdot\hat{\bm{r}}} \hat{P}_{\bm{k}\mathcal{B}}\bigg(\partial_{k_i}\hat{Q}_{\bm{k}\mathcal{B}}\bigg)\bigg[\bigg(\partial_{k_{j}}\partial_{k_{k}} \hat{P}_{\bm{k}\mathcal{B}}\bigg)\hat{P}_{\bm{k}\mathcal{B}}e^{-i\bm{k}\cdot\hat{\bm{r}}} +\bigg(\partial_{k_{j}}\hat{P}_{\bm{k}\mathcal{B}}\bigg)\bigg(\partial_{k_{k}} \hat{P}_{\bm{k}\mathcal{B}}e^{-i\bm{k}\cdot\hat{\bm{r}}} \bigg)+\bigg(\partial_{k_{k}} \hat{P}_{\bm{k}\mathcal{B}}\bigg)\bigg(\partial_{k_{j}}\hat{P}_{\bm{k}\mathcal{B}}e^{-i\bm{k}\cdot\hat{\bm{r}}} \bigg)\bigg]\\
    &=\hat{Q}^{(3)}_{ijk}+\hat{Q}^{(2)}_{ij}\hat{r}_{k\mathcal{B}}+\hat{Q}^{(2)}_{ik}\hat{r}_{j\mathcal{B}},
\end{align}
 where we introduced the geometric quantity $\hat{Q}^{(3)}_{ijk}=\sum_{n\in\mathcal{B}}\sum_{\bm{k}}{Q}^{(3)}_{ijk,nm}(\bm{k})\ket{\psi_{n\bm{k}}}\bra{\psi_{m\bm{k}}}$, whose matrix elements are\cite{Kozii2021} ${Q}^{(3)}_{ijk,nm}(\bm{k})=i\bra{u_{n\bm{k}}}\left(\partial_{k_i}\hat{P}_{\bm{k}\mathcal{B}}\right)\left( \partial_{k_{j}}\partial_{k_{k}} \hat{P}_{\bm{k}\mathcal{B}}\right)\ket{u_{m\bm{k}}} $. Its relation to quantum Christoffel symbols are given by eq.(\ref{eq:q3}).

\section{Canonical momentum }
\subsection{Canonical momentum and and real space Berry connections}
The canonical momentum is defined as the operator such that 
\begin{align}
    \bra{\bm{r}}\hat{\bm p}\ket{\psi}
    =
    -i\partial_{\bm r} \psi(\bm{r}).
\end{align}
In Bloch basis we may decompose it to be\cite{Blount1962FormalismsOB} 
$
  \hat{\bm{p}}  
    =
    \hat{\bm{k}}- \hat{\mathcal{A}}^{\bm{r}}, 
$
with matrix elements 
\begin{align}
     &\bra{\psi_{n\bm{q}}}\hat{\bm{p}}  \ket{\psi_{m\bm{k}}}
    =
    \left(
    \bm{k} \delta_{nm}
    -
    \mathcal{A}^{\bm{r}}_{nm}(\bm{q})
    \right)
    \delta(\bm{q-k}),
\end{align}
where the second term is a real space Berry connection 
\begin{align}
    &\mathcal{A}^{\bm{r}}_{nm}(\bm{q})
    =
    i\bra {u_{n\bm{q}}}\partial_{\bm{r}}\ket{u_{m\bm{q}}} 
   \equiv 
    i\frac{(2\pi)^d}{V_{uc}} \int_{cell} d\bm{r}   u^*_{n\bm{q}}(\bm r) \partial_{\bm{r}}u_{m\bm{q}}(\bm r)  .
\end{align}  
It may be seen from transforming to Wannier basis that the Bloch momentum operator generates inter-cell translations, while the real space Berry connection generates intra-cell displacements. We would like to point out that this real space Berry connection is different from the Berry connection that results from position dependent Bloch states in a spatially varying band structure.

\subsection{Mixed geometrical quantities}
The coupling of the canonical momentum and vector potential in the presence of magnetic field induces contributions that are expressed in terms of new geometric quantities, which depends on the spatial gradient of the cell-periodic part of Bloch states as well. 
\subsubsection{Mixed quantum geometric tensor}
First we introduce a mixed quantum geometric tensor defined as
\begin{align}
    \hat{\mathcal L}_{ij}%
    \equiv
    -\hat{P}_{\mathcal{B}}
    [\hat{r}_{i  },\hat{Q}_{\mathcal{B}}]
    [\hat{p}_{j  },\hat{Q}_{\mathcal{B}}]
    \hat{P}_{\mathcal{B}}
    =
    ((\hat{r}_i-\hat{r}_{i\mathcal{B}})
    \hat{p}_{j  })_{\mathcal{B}}\label{eq.mixedl1}.
\end{align}
We can see from the definition that $\hat{\mathcal{L}}_{ij}$ is odd under time reversal transformation and even under inversion. For any Bloch diagonal operator $\hat{O}$ we may define a local operator
\begin{align}
    \hat{O}_{\bm{q}}
    \equiv
    e^{-i\bm{q\cdot\hat{r}}} 
     \hat{O}
     e^{i\bm{q\cdot\hat{r}}} ,\label{local}
\end{align}
whose matrix elements of the local operator evaluated on cell-periodic Bloch functions are the matrix elements of $\hat{O}$ in Bloch states, namely 
\begin{align}
\bra{u_{n\bm{q}}}\hat{O}_{\bm{q}}\ket{u_{m\bm{q}}}
   =
    {O}_{nm}(\bm{q})
    =
    \bra{\psi_{n\bm{q}}}\hat{O} \ket{\psi_{m\bm{q}}}.
\end{align}
Now we compute
\begin{align}
      \hat{\mathcal L}_{ij}
    &=
    \lim_{\bm{a}\rightarrow 0} \ri\partial_{a_i} 
    \hat{P}_N 
    e^{-i\bm{a\cdot\hat{r}}}
    \hat{Q}_N 
    \hat{ p}_j 
 \hat{P}_N    
= \ri e^{i\bm{q\cdot\hat{r}}}
 \hat{P}_{\bm{q}\mathcal{B}}
 \left
 (\partial_{q_i}
 \hat{Q}_{\bm{q}\mathcal{B}} 
 \right)   
 \hat{Q}_{\bm{q}\mathcal{B}} 
 \hat{A}^{\bm{r}}_{j\bm{q}}
 \hat{P}_{\bm{q}\mathcal{B}}
 e^{-i\bm{q\cdot\hat{r}}} ,
\end{align}
where in the last step we used that $\hat{k}$ commutes with projection operator. That is we have 
\begin{align}
    \hat{\mathcal L}_{ij,nm}(\bm{q})
    \delta(\bm{q}-\bm{k}) 
    =
    \bra{\psi_{n\bm{q}}}
    \hat{\mathcal L}_{ij }
    \ket{\psi_{m\bm{k}}}
    =
    \ri \bra{u_{n\bm{q}}}
    \left
 (\partial_{q_i}
 \hat{Q}_{\bm{q}\mathcal{B}} 
 \right)   
 \hat{Q}_{\bm{q}\mathcal{B}} 
 \hat{A}^{\bm{r}}_{j\bm{q}}
    \ket{u_{m\bm{q}}}\delta(\bm{q}-\bm{k}).
\end{align}
This is a new geometric quantity that is local in momentum space and depends on the spatial gradient of the Bloch states, due to $\hat{A}^{\bm{r}}_{j\bm{q}}$, in addition to momentum gradient. 
\subsubsection{Modified canonical commutation relation }
The commutation relation between the projected position and canonical momentum becomes
\begin{align}
    [\hat{  r}_{ i\mathcal{B}},\hat{  p}_{ j\mathcal{B}}]
    &=
    \ri\mathbb{1}_{\B} \delta_{ij} - \ri[\hat{\nabla}_i,\hat{ {\mathcal A}}^{\bm{r}}_{j\mathcal{B}}]\label{eq:modCCM}
    .
\end{align} 
This shows that the projected position and momentum no longer form a canonical pair. But note that we also have
\begin{align}
    [\hat{  r}_{ i\mathcal{B}},\hat{  p}_{ j\mathcal{B}}]
    &=
    [\hat{  r}_{ i },\hat{  p}_{ j} ]_{  \mathcal{B}}
    -
    \bigg( 
    \hat{  r}_{i}\hat{Q}_N \hat{  p}_{j}-\hat{  p}_{j}\hat{Q}_N \hat{  r}_{i}   \bigg)_{  \mathcal{B}}
    \\
    &=
     \ri\hat{P}_\mathcal{B}\delta_{ij}
    -
    \bigg( 
    \hat{\mathcal{L}}_{ij}-\hat{\mathcal{L}}_{ij}^\dagger 
    \bigg)
    ,
\end{align}
from which we have that 
\begin{align}
    [\hat{\nabla}_i,\hat{ {\mathcal A}}^{\bm{r}}_{j\mathcal{B}}]=\frac{1}{ \ri}\bigg( \hat{\mathcal{L}}_{ij}-\hat{\mathcal{L}}_{ij}^\dagger \bigg),
\end{align}
which is manifestly hermitian.

\subsubsection{Higher mixed geometric quantities}
Consider now the quantity $ \proj{\hat{r}_i\hat{r}_j\QB\hat{p}_k}$. We have that 
\begin{align}
    &\proj{\hat{r}_i\hat{r}_j\QB\hat{p}_k}
    =
    \hat{r}_{i\mathcal{B}} \proj{\hat{r}_j\QB\hat{p}_k}
    +
     \proj{\hat{r}_i\QB\hat{r}_j\QB\hat{p}_k}
     \equiv
     \ri \hat{\nabla}_i\hat{\mathcal{L}}_{jk}     %
     +
     \hat{\Lambda}_{ijk} ,\label{eq:mixed rrp}
\end{align}
where $\hat{\Lambda}_{ijk}\equiv\proj{\hat{r}_i\QB\hat{r}_j\QB\hat{p}_k}= \proj{[\hat{r}_i,\QB]\hat{r}_j\QB[\hat{p}_k,\PB]}$, and we used $\hat{r}_{i\mathcal{B}}=\ri\hat{\nabla}_i$.  Note that by symmetry we have 
\begin{align}
    \hat{\Lambda}_{ijk}-\hat{\Lambda}_{jik}=-\ri\left( \hat{\nabla}_i\hat{\mathcal{L}}_{jk}-\hat{\nabla}_j\hat{\mathcal{L}}_{ik}\right).
\end{align} 

In fact we have 
\begin{align}
    \hat{\Lambda}_{ijk}
    &
    =
    \proj{[\hat{r}_i,\QB]\hat{r}_j\QB[\hat{p}_k,\PB]}
    \\&=
    \proj{[[\hat{r}_i,\QB],\hat{r}_j]\QB[\hat{p}_k,\PB]}
    +
    \proj{[\hat{r}_i,\QB]\QB\hat{r}_j\QB[\hat{p}_k,\PB]}
    \\&=
     \proj{[\hat{r}_j,[\hat{r}_i,\PB] ]\QB[\hat{p}_k,\PB]}
    -
    \proj{ \hat{r}_j \PB[\hat{r}_i,\PB]\QB[\hat{p}_k,\PB]}
    \\&=
    \hat{\mathcal{L}}_{ijk}^{(2)}+i\hat{\nabla}_j\hat{\mathcal{L}}_{ik},
\end{align}
where we introduced a geometric quantity $\hat{\mathcal{L}}_{ijk}^{(2)}\equiv \proj{[\hat{r}_j,[\hat{r}_i,\PB]]\QB[\hat{p}_k,\PB]}$, which is symmetric in $ij$ and local in momentum space. Substituting this back in \cref{eq:mixed rrp}, we have
 \begin{align}
\proj{\hat{r}_i\hat{r}_j\QB\hat{p}_k}=\hat{\mathcal{L}}_{ijk}^{(2)}+i\hat{\nabla}_i\hat{\mathcal{L}}_{jk}+i\hat{\nabla}_j\hat{\mathcal{L}}_{ik},
 \end{align}
which is manifestly $ij$ symmetric as it should.

%===========================================================================
\section{Equations of motion}
%===========================================================================
In this section, we derive the operator dynamics presented in \cref{eq:constrained_dynamics_r,eq:constrained_dynamics_k2} in the Heisenberg picture.
In the below assume throughout that the operators do not carry explicit time-dependence, in particular $\partial\hat O/\partial t = 0$.

%===========================================================================
\subsection{Generality}
\subsubsection{Velocity}
%===========================================================================
The velocity is obtained by direct evaluation of the projected dynamics of the band-projected position operator,
\begin{align}
    \PB \left(\frac{\rd}{\rd t}\RB\right) \PB
    =
    -\ri \left[\RB, \HB\right]
    =
    -\ri \left[\R, \H\right]_{\mathcal B}
    +
    \ri\left(
        \R \QB \H - \H \QB \R
    \right)_{\mathcal B}
    .
    \label{eq:constrained_dynamics_r_details}
\end{align}
Note that we have $\H = \Hbloch + \Hpert$, in which the Bloch Hamiltonian $\Hbloch = \sum_{n\bm k}\E_{n\bm k}\ket{\psi_{n\bm k}}\bra{\psi_{n\bm k}}$ is a band-diagonal operator, therefore $[\QB, \Hbloch] = 0$, such that the last bracket in \cref{eq:constrained_dynamics_r_details} equals the inter-band velocity $\dtRinterB = \ri\left(\R \QB \Hpert - \Hpert \QB \R\right)_{\mathcal B}$.
The remaining term is the intra-band velocity
\begin{align}
    -\ri\left[\R, \H\right]_{\mathcal B}
    =
    -\ri \left[\R, \hat{\bm\pi}^2/(2m)\right]_{\mathcal B}
    -\ri \left[\R, q\phi(\hat{\bm r}) + V(\hat{\bm r})\right]_{\mathcal B}
    =
    -\ri \frac1{2m}
    \left\{
        \left[\R, \CanMc{i} - q A_i(\R)\right]
        ,
        \hat\pi_i
    \right\}_{\B}
    =
    \frac{\hat{\bm \pi}_{ \B}}{m}
\end{align}
which concludes the derivation of  
\cref{eq:constrained_dynamics_r}. In the below, we will use gradient expansion to discuss the inter-band velocity terms that give rise to corrections to the equation of motion. 

%===========================================================================
\subsubsection{Bloch momentum dynamics}
%===========================================================================
Note any band-diagonal operator commutes with the projection operator, hence, in particular, $\PB  \hat{\bm{k}} \PB=\PB\hat{\bm{k}}  = \hat{\bm{k}} \PB$.
 Therefore, the projected dynamics of the projected crystal momentum is equivalent to the projected dynamics of the crystal momentum, i.e. 
\begin{align}
    \ri\proj{\dot{\hat{k}}_{i\mathcal{B}}}
    =
    [\PB\hat{k}_i \PB,\PB\hat{H}\PB]
    =
    \PB[\hat{k}_{i},\hat{H}]\PB
    ,
\end{align}
where $\H = \frac{\hat{\bm \pi}^2}{2m} + q\phi(\hat{\bm r}) + V(\hat{\bm r})$.
Using  $[\hat{k}_{i},f(\hat{\bm{r}})]=-i\left[\partial_{\hat r_i},f(\hat{\bm{r}})\right]$ we have, in Coulomb gauge,
\begin{align}
     \dot{\hat{k}}_{i\mathcal{B}} 
    %1
    &=
    %1
    \frac{1}{\ri}\hat{P}_{\mathcal{B}}[\hat{k}_i  ,  U(\hat{\bm p},\hat{\bm r})]\hat{P}_{\mathcal{B}}
    %2
    \\
    &=
    %2
    \frac{1}{\ri}\hat{P}_{\mathcal{B}}[\hat{k}_i  ,  q\phi(\hat{\bm r})-\frac{q\bm{A}(\bm {\hat{r}})\cdot\bm{p }}{m}+\frac{q^2\bm{A }^2(\hat{\bm r})}{2m}]\hat{P}_{\mathcal{B}}
    %3
    \\
    &=
    %3
    -q \left(  \partial_{r_i}\phi(\hat{\bm{r}})   \right)_{\mathcal{B}}+\frac{q}{m}   \left( \left( \partial_{r_{i}}  \bm{A}\right)   \cdot (\hat{\bm{p}}-q\bm{A})  \right)_{\mathcal{B}}
    %4
    \\
    &=
    %4
    qE_{i\mathcal{B}}(\hat{\bm r})+\frac{q}{m}\left(\partial_{r_{i}}  \bm{A}   \cdot \bm{\hat{\pi}}\right)_{\mathcal{B}}
    %5
    \\
    & =              
    %5
    qE_{i\mathcal B}(\R) + \frac q m\left(\left[\partial_{\hat r_i}, A_j(\R)\right]\hat\pi_j\right)_{\mathcal B}
\end{align}

On the other hand, to first order in the applied fields, the projected dynamics of the gauge-invariant crystal momenta   $\hat{\bm{\kappa}}=\hat{\bm{k}}-q\bm{A}(\hat{\bm{r}})$  can be approximately taken as 
\begin{align} 
    \frac{d}{dt}\hat{\bm{\kappa}}_{\mathcal{B}}
    \approx
     q\bm{E}_{ \mathcal B}(\R)
     +
    \frac{q}{2m} 
     \bigg(
     \bm{\hat{\pi}} \times  \bm{B}(\hat{\bm{r}})
     -
     \bm{B}(\hat{\bm{r}})\times\bm{\hat{\pi}} 
     \bigg)_{\mathcal{B}}
      .
\end{align}
 This holds in any gauge that are static, which we can always assume to exist for static fields. Note that it is the gauge-invariant $\bm{\pi}$ and not $\dot{\bm {\hat{r}}}$ that appears in the force equation.

\subsection{Position dynamics in electric field}
\subsubsection{Velocity}
In an electric field the only intra-band contribution is due to the Bloch Hamiltonian $\Hbloch$, and is just the projected velocity operator $\hat{\bm{v}}_{\mathcal{B}}$. The inter-band velocity in an electric field is 
\begin{align}
    \dot{\hat{r}}_{i\perp}^{\phi}
    &=
    -\frac{q}{\ri} 
    \proj{
        \hat{r}_{i } \QB \phi(\R) - \phi(\R) \QB \hat{r}_{i }}
    =
    \frac{q}{\ri}
   [ \hat{r}_{i \mathcal{B}}, \phi_{\mathcal{B}}(\R) ]
\end{align}
Expanding the scalar potential as 
\begin{align}
    \phi(\R)
=
-E^{(0)}_i\hat{r}_i
-\frac{1}{2!} 
  E^{(1)}_{ij }\hat{r}_i\hat{r}_j
-
\frac{1}{3!}  
 E^{(2)}_{ijk}\hat{r}_i\hat{r}_j\hat{r}_k
 ,
 \end{align}
 then this can be dealt with the method outlined in Section I. This gives 
 \begin{align}
        \dot{\hat{r}}_{i\perp}^\phi
        =
 -\frac{q}{2}\{\hat{\Omega}_{ij}, E_{j,\mathcal B}(\R)\}
    +
     \hat R_{i\perp}^\phi
    ,
 \end{align}
where to second order expansion 
\begin{align}
    \hat R_{i\perp}^\phi=-\frac{q}{2}E_{jk}^{(2)}[\hat{\nabla}_i,\hat{\mathcal{G}}_{jk}] 
\end{align}

\subsubsection{Anomalous velocity}
Here we illustrate the emergence of anomalous velocity by an explicit calculation. First note that for a uniform electric field $\phi (\R)=   - E^{(0)}_{i  }\hat{r}_i  $, we have $ \ri\dot{\hat{r}}_{i\perp }^\phi =- qE^{(0)}_{j  }[\hat{r}_{i\B}, \hat{r}_j]=-i  qE^{(0)}_{j  }\hat{\Omega}_{ij}$. Now consider the lowest order non-uniform electric field given by the scalar potential   $\phi (\R)=   -\frac{1}{2!}E^{(1)}_{ij }\hat{r}_i\hat{r}_j $, or the electric field being $E_i(\R)= E^{(1)}_{ij } \hat{r}_j $.
\\
\\
We have that, recalling eq.(\ref{eq:2nd_geometric_operators}) 
\begin{align}
    [\hat{r}_{i\B},\proj{\hat{r}_j\hat{r}_k}]&
    =\frac{1}{2} [\hat{r}_{i\B},\proj{\{\hat{r}_j,\hat{r}_k\}}]
    \\&=\frac{1}{2} [\hat{r}_{i\B}, \{\hat{r}_{j\B},\hat{r}_{k\B}\} ]+\frac{1}{2} [\hat{r}_{i\B}, 2\hat{\mathcal G}_{jk} ]\\
     \\&=\frac{1}{2}\ri\{\hat{\Omega}_{ij},\hat{r}_{k\B} \}+\frac{1}{2}\ri\{\hat{\Omega}_{ik},\hat{r}_{j\B} \}+  [\hat{r}_{i\B},  \hat{\mathcal G}_{jk} ]
\end{align}
Thus we have
\begin{align}
    i\hat r_{i\perp}^\phi&=[\hat{r}_{i\B},q\phi_\B(\hat{\bm{r}})]
    =-\frac{q}{2 }E^{(1)}_{jk } [\hat{r}_{i\B},\proj{\hat{r}_j\hat{r}_k}]
    \\&=-\frac{q}{2 }E^{(1)}_{jk } \bigg(
    \frac{1}{2}\ri\{\hat{\Omega}_{ij},\hat{r}_{k\B} \}+\frac{1}{2}\ri\{\hat{\Omega}_{ik},\hat{r}_{j\B} \}+  [\hat{r}_{i\B},  \hat{\mathcal G}_{jk} ]\bigg)
     \\&=-\ri\frac{q}{4 }  \bigg(
     \{\hat{\Omega}_{ij},E_{j\B}(\R) \}+  \{\hat{\Omega}_{ik},E_{k\B}(\R) \}\bigg)   -\frac{q}{2 }E^{(1)}_{jk }[\hat{r}_{i\B},  \hat{\mathcal G}_{jk} ]
     \\&=-\ri\frac{q}{2 }   
     \{\hat{\Omega}_{ij},E_{j\B}(\R) \}    -\frac{q}{2 }E^{(1)}_{jk }[\hat{r}_{i\B},  \hat{\mathcal G}_{jk} ]
\end{align}
or 
\begin{align}
   \hat r_{i\perp}^\phi=- \frac{q}{2 }   
     \{\hat{\Omega}_{ij},E_{j\B}(\R) \}    -\frac{q}{2 }E^{(1)}_{jk }[\hat{\nabla}_{i },  \hat{\mathcal G}_{jk} ] 
\end{align}
This is the operator form of the results obtained in \cite{Lapa2019} and \cite{Kozii2021}, generalized to the multi-band case. It is clear that the pattern of the decomposition of the inter-band velocity into the sum of a Berry curvature induced anomalous velocity and additional geometric terms continue in higher orders of inhomogeneity.

\subsection{Dynamics in magnetic fields}
  \subsubsection{General potential}
The Hamiltonian in a magnetic field reads $\hat{H}= \bm{\hat{\pi}} ^2/{2m}+V(\hat{\bm{r}})=\Hbloch+ \Hpertdia + \Hpertpara $, where $\Hpertdia = q^2 \bm A(\R)^2/(2m$) and $\Hpertpara = -q (\bm A(\R) \cdot \CanM+\CanM\cdot\bm A(\R))/(2m)$. The
projected dynamics of the position operator is given by
\begin{align}
    \ \frac{\rd}{\rd t}\RB   
    &=
    -\ri \left[\RB, \HB\right]
    =
   \frac{\hat{\bm \pi}_{\mathcal B}}{m}
    +
    \ri\proj{
        \R \QB \Hpert - \Hpert \QB \R}\\
    &=
   \frac{\hat{\bm \pi}_{\mathcal B}}{m}
    +  
    \dtRinterB
    ,
\end{align}
where we used the fact that $\proj{\QB\Hbloch}=0$ to obtain the second equality. The first intra-band term is the gauge-invariant kinematic momentum in ordinary quantum mechanics. The inter-band velocity due to the diamagnetic term reads 
\begin{align}
    \ri \dot{\hat{r}}_{j\perp}&=\frac{q^2}{2m}
     [\hat{r}_{j\B}, \proj{A_k (\R)A_k (\R) } ] 
    \\&=\frac{q^2}{2m}
    \left(
    \left\{
     [\hat{r}_{j\B}, A_{k\B}(\R) ], A_{k\B}(\R)\right\}+ [\hat{r}_{j\B}, \proj{A_{k}(\R)\QB A_{k}(\R)}]  
    \right)
\end{align}
To compute the paramagnetic contribution it is easier to consider first the whole paramagnetic term first, including the intra-band velocity. Namely 
\begin{align}
     \ri\dot{\hat r}_{j\perp}&=[\hat {r}_{j\mathcal{B}},\Hpertpara]
     =
     -\frac{ q}{2m} [\hat {r}_{j\mathcal{B}}  ,  
   \proj{\{
     {A}_{k}(\bm {\hat{r}}), \hat p_{k} 
   \}} ]
    \\&=
     -\frac{ q}{2m} \left(
     [\hat {r}_{j\mathcal{B}}  ,  
   \{
     {A}_{k\mathcal{B}}(\bm {\hat{r}}) , \hat p_{k  \mathcal{B}}
   \}  ]
   +[
   \hat {r}_{j\mathcal{B}}  ,  
   \proj{
     {A}_{k }(\bm {\hat{r}}) \QB \hat p_{k } + \hat p_{k }\QB  {A}_{k }(\bm {\hat{r}}) 
     }
      ]
      \right)
    \\&=
     -\frac{q}{2m}\left( 
     \{ [\hat {r}_{j\mathcal{B}}  ,  
  \hat p_{k  \mathcal{B}}   
    ],  {A}_{k\mathcal{B}}(\bm {\hat{r}}) \}
    +
      \{ [\hat {r}_{j\mathcal{B}}  ,  
     {A}_{k\mathcal{B}}(\bm {\hat{r}}) 
    ], \hat p_{k  \mathcal{B}} \}
   +
   [
   \hat {r}_{j\mathcal{B}}  ,  
   \proj{
     {A}_{k }(\bm {\hat{r}}) \QB \hat p_{k  }
    +   \hat p_{k  }\QB{A}_{k }(\bm {\hat{r}})  } ]
    \right)
\end{align}
Now we subtract the intra-band contribution due to the paramagnetic perturbation which is $-\frac{q}{m}\ri {A}_{j\mathcal{B}}(\bm {\hat{r}}) $ from the first term in the bracket,  and using Eq.(\ref{eq:modCCM}), we obtain
\begin{align}
    \ri\dot{\hat r}_{j\perp} &=
     -\frac{q}{2m}\left( -
     \{ [\hat {r}_{j\mathcal{B}}  ,  
  \hat{ {\mathcal A}}^{\bm{r}}_{k\mathcal{B}}
    ],  {A}_{k\mathcal{B}}(\bm {\hat{r}}) \}
    +
      \{ [\hat {r}_{j\mathcal{B}}  ,  
     {A}_{k\mathcal{B}}(\bm {\hat{r}}) 
    ], \hat p_{k  \mathcal{B}} \}
   +
   [
   \hat {r}_{j\mathcal{B}}  ,  
   \proj{
     {A}_{k}(\bm {\hat{r}}) \QB \hat p_{k  }
    +  \hat p_{k  }\QB{A}_{k }(\bm {\hat{r}})  } ]
    \right)
\end{align}
The first term emerges because the projected position and momentum are no longer a canonical pair but rather satisfies Eq.(\ref{eq:modCCM}). The second term arises as the projected position operators no longer commute with each other, and can be dealt with the same method as the electric field case, whereas the last one introduces new, mixed geometric quantities that involve spatial gradient of Bloch states as well. 
\\
\\
Finally combining the diamagnetic and paramagnetic terms we obtain \cref{eq:magfield_1,eq:magfield_2,eq:magfield_3,eq:magfield_4} in the main text
\begin{align}
    \ri\dot{\hat r}_{j\perp} =&
     -\frac{q}{2m}\left( -
     \{ [\hat {r}_{j\mathcal{B}}  ,  
  \hat{ {\mathcal A}}^{\bm{r}}_{k\mathcal{B}}
    ],  {A}_{k\mathcal{B}}(\bm {\hat{r}}) \}
    +
      \{ [\hat {r}_{j\mathcal{B}}  ,  
     {A}_{k\mathcal{B}}(\bm {\hat{r}}) 
    ],  \hat{\pi}_{k  \mathcal{B}} \}
   +
   [
   \hat {r}_{j\mathcal{B}}  ,  
   \proj{
     {A}_{k}(\bm {\hat{r}}) \QB  p_{k  }
    +   p_{k  }\QB{A}_{k }(\bm {\hat{r}})  } ]
    \right)
\ \\&+\frac{q^2}{2m}
    \left(
      [\hat{r}_{j\B}, \proj{A_{k}(\R)\QB A_{k}(\R)}]  
    \right)\label{eq:magnetic eom}
\end{align}

\subsubsection{Gradient expansion}
So far the vector potential has been assumed to be geneal. We now expand the magnetic vector potential as $A_i(\hat{\bm{r}})= A^{(1)}_{ij}\hat{r}_j+\frac{1}{2}A^{(2)}_{ijk}\hat{r}_j\hat{r}_k$, we have the following respective geometric contributions to the inter-velocity. 
\\
\\
\textbf{First order}
To linear order in the gradient expansion gives
\begin{align}
     \dot{\hat{r}}^{(1)}_{l\perp } 
    =& \
    \ri A_{jk}^{(1)}\frac{q}{m}\bigg(\{ \hat{\nabla}_{k } , [\hat{\nabla}_l,\hat{ {\mathcal A}}^{\bm{r}}_{j\mathcal{B}}] \}
     +
     \{ \hat{\Omega}_{lk  },\hat{\pi}_{j \mathcal{B}}\}
     -
     [\hat{\nabla}_{l },   \hat{\mathcal L}_{jk}] 
    \bigg)
   \\&\ +  \frac{q^2}{2m}A^{(1)}_{ka}A^{(1)}_{kb}[\hat{\nabla}_l,\hat{\mathcal Q}^{(2)}_{ab}]
    .
\end{align}
Since $a$ and $b$ are dummy variables, the last term due to diamagnetic term can be re-written as 
\begin{align}
    \frac{q^2}{2m}A^{(1)}_{ka}A^{(1)}_{kb}[\hat{\nabla}_l,\hat{\mathcal Q}^{(2)}_{ab}]= \frac{q^2}{2m}A^{(1)}_{ka}A^{(1)}_{kb}[\hat{\nabla}_l,\hat{\mathcal G}^{(2)}_{ab}],
\end{align}
which is manifestly Hermitian.
\\
\\
\textbf{Second order} 
To second order in the gradient expansion, we have
\begin{align}
    \dot{\hat r}_{i\perp}^{(2)}
    &=
     \frac{+q}{2m}A^{(2)}_{jkl} \{   [\hat{\nabla}_i,\hat{ {\mathcal A}}^{\bm{r}}_{j\mathcal{B}}],  \proj{\hat{r}_k\hat{r}_l} \}
    -
     \frac{ q}{2m} A^{(2)}_{jkl}\{ [\hat {\nabla}_{i\mathcal{B}}  ,  
    \proj{\hat{r}_k\hat{r}_l} 
    ],  p_{j  \mathcal{B}} \}
   \\&-
     \frac{q}{2m}A^{(2)}_{jkl}\bigg(
   [
   \hat {\nabla}_{i }  ,  
    \ri(\hat{\mathcal{L}}_{lj}+\hat{\mathcal{L}}_{lj}^\dagger)\hat{\nabla}_k+\ri(\hat{\mathcal{L}}_{kj}+\hat{\mathcal{L}}_{kj}^\dagger)\hat{\nabla}_l+i[\hat{\nabla}_k,\hat{\mathcal{L}}_{lj}] +i[\hat{\nabla}_l,\hat{\mathcal{L}}_{kj}] ]
  +
   [
   \hat {\nabla}_{i }  ,  
    \hat{\mathcal{L}}^{(2)}_{klj}+\hat{\mathcal{L}}^{(2) \dagger}_{klj} ]\bigg),
\end{align}
where in arriving at the second line we used
\begin{align}
     \ri \hat{\nabla}_k\hat{\mathcal{L}}_{lj}+ \ri \hat{\mathcal{L}}_{lj}^\dagger\hat{\nabla}_k  
       =
      i(\hat{\mathcal{L}}_{lj}+\hat{\mathcal{L}}_{lj}^\dagger)\hat{\nabla}_k+i[\hat{\nabla}_k,\hat{\mathcal{L}}_{lj}].
\end{align}
Note that using Eq.(\ref{eq:2nd_geometric_operators}), we have 
\begin{align}
     \proj{\hat{r}_k\hat{r}_l}= \hat r^{(2)}_{kl,\mathcal B} 
    =
    \ri^2
    \left(
        \frac12\{\hat\nabla_k,\hat\nabla_l\}
        -
        \hat{\mathcal G}_{kl}
    \right)
    ,
\end{align}
where $\hat{\mathcal G}_{kl}$ is the quantum metric tensor.

\end{document}